\documentclass[fleqn,usenatbib]{mnras}

\usepackage{mathptmx}
\usepackage{threeparttable}

\usepackage[T1]{fontenc}
\usepackage{ae,aecompl}

\usepackage{graphicx} 
\usepackage{amsmath} 
\newcommand{\angstrom}{\textup{\AA}}
\usepackage{amssymb} 
\usepackage{amsmath}
\usepackage[utf8]{inputenc}
\usepackage{mathptmx}
\usepackage{makecell}
\usepackage{appendix}

\usepackage{footnote}
\makesavenoteenv{tabular}
\makesavenoteenv{table}

\title[Rapid Optical/X-ray correlations of MAXI\,J1820+070]{The Evolution of Rapid Optical/X-ray Timing Correlations in the Initial Hard State of MAXI\,J1820+070}

\author[J. A. Paice et al.]{J. A. Paice$^{1,2}$\thanks{E-mail: j.a.paice@soton.ac.uk},
	P. Gandhi$^{1}$,
	T. Shahbaz$^{3,4}$,
	A. Veledina$^{5,6}$,
	J. Malzac$^{7}$,\newauthor
	D.A.H. Buckley$^{8,9}$,
	P. A. Charles$^{1}$,
	K. Rajwade$^{10}$,
	V. S. Dhillon$^{11,3}$,\newauthor
	S. P. Littlefair$^{11}$,
	T. R. Marsh$^{12}$,
	P. Uttley$^{13}$,
	F. M. Vincentelli$^{1}$,
	and R. Misra$^{2}$.
	\\
	$^{1}$Department of Physics and Astronomy, University of Southampton, Highfield, Southampton, SO17 1BJ, UK\\
	$^{2}$Inter-University Centre for Astronomy and Astrophysics, Pune, Maharashtra 411007, India\\
	$^3$Instituto de Astrof\'\i{}sica de Canarias (IAC), E-38205 La Laguna,  Tenerife, Spain \\
    $^4$Departamento de  Astrof\'\i{}sica, Universidad de La Laguna (ULL),  E-38206 La Laguna, Tenerife, Spain \\
    $^{5}$Department of Physics and Astronomy, FI-20014 University of Turku, Finland\\
	$^{6}$Nordita, KTH Royal Institute of Technology and Stockholm University, Roslagstullsbacken 23, SE-10691 Stockholm, Sweden\\
	$^{7}$IRAP Universite de Toulouse, CNRS, UPS, CNES, Toulouse, France\\ 
	$^{8}$South African Astronomical Observatory, Observatory Road, Observatory, 7925, Cape Town, South Africa\\ 
    $^{9}$Department of Astronomy, University of Cape Town, Private Bag X3, Rondebosch 7701, South Africa \\ 
    $^{10}$Department of Physics and Astronomy, The University of Manchester, Oxford Road, Manchester, M13 9PL, UK\\
	$^{11}$Department of Physics and Astronomy, University of Sheffield, Sheffield, S3 7RH, UK\\ 
	$^{12}$Astronomy and Astrophysics Group, Department of Physics, University of Warwick, Gibbet Hill Road, Coventry, CV4 7AL, UK\\ 
	$^{13}$Astronomical Institute 'Anton Pannekoek', University of Amsterdam, Science Park 904, NL-1098XH Amsterdam, the Netherlands
}

\date{Submitted to MNRAS in original form XXXX AAA XX, Received XXXX AAA XX, Accepted XXXX AAA XX}

\pubyear{2019}

\begin{document}
	\label{firstpage}
	\pagerange{\pageref{firstpage}--\pageref{lastpage}}
	\maketitle
	
	\begin{abstract}
	{We report on a multi-epoch campaign of rapid optical/X-ray timing observations of the superbright 2018 outburst of MAXI J1820+070, a black hole low-mass X-ray binary system. The observations spanned 80 days in the initial hard-state, and were taken with NTT/ULTRACAM and GTC/HiPERCAM in the optical ($u_sg_sr_si_sz_s$ filters at time resolutions of 8--300\,Hz) and with ISS/NICER in X-rays. We find (i) a growing anti-correlation between the optical and X-ray lightcurves, (ii) a steady, positive correlation at an optical lag of $\sim$0.2\,s (with a longer lag at longer wavelengths) present in all epochs, and (iii) a curious positive correlation at \textit{negative} optical lags in the last, X-ray softest epoch, with longer wavelengths showing a greater correlation and a more negative lag. To explain these we postulate the possible existence of two synchrotron-emitting components; a compact jet and a hot flow. In our model, the significance of the jet decreases over the outburst, while the hot flow remains static (thus, relatively, increasing in significance). We also discuss a previously discovered quasi-periodic oscillation and note how it creates coherent optical time lags, stronger at longer wavelengths, during at least two epochs.}	
		
	\end{abstract}
	

	
	\section{Introduction} \label{sec:Intro}

    

    Low-Mass X-ray Binaries (LMXBs) are highly variable systems involving accretion on to either a neutron star or black hole. Over the past few decades, there have been many efforts to study this variability and detail its behaviour, and there is an expanding body of literature detailing this (see \citealt{BelloniVariability1990, Mushotzky_X-ray_1993, Klis_Millisecond_2000} and many others). But why is the study of this variability important?
    
    In short, because these systems are complex, and unresolvable with current telescopes. LMXBs host a compact object accreting via a disc of material transferred from a Roche-lobe-filling companion star. The environment is complex, with an outer disc, hot inner flow/corona, and compact, relativistic jets (to name just a few), which all emit across the electromagnetic spectrum. And, during transient, violent outbursts that occur every few years or decades and can last for weeks to months, the scale, presence, and behaviour of these regions can change significantly. Their compact nature means that physically important timescales can span $\sim$microseconds in the inner zones, to decades at the other extreme. The goal of multiwavelength timing studies of these sources is to understand the physical processes in these components, and thus the system as a whole.
    
    Over the relatively short history of multiwavelength astronomy, better technology and new telescopes have improved the temporal resolution of such studies, and with it, our understanding has advanced; \citet{Motch_GX339_1982, motch_simultaneous_1983} and \cite{Imamura_1.13msGX339_1987} were some of the earliest reports on rapid stochastic multiwavelength variability down to millisecond scales, while \citet{kanbach_correlated_2001} was one of the works that showed intriguing relations between the rapid optical and X-ray variability for the first time. This interband relationship was then found to vary between systems, each time showing complex behaviour, interpreted as a varying dominance of the inflowing or outflowing plasma through the disc, the inner flow and the jet (e.g. \citealt{Gandhi_Correlations_2008, durant_swift_2008, gandhi_rapid_2010, casella_fast_2010, Durant_OptXCCFs_2011, gandhi_elevation_2017, Pahari_BWCir_2017}). The true importance of these studies is in providing novel quantitative constraints of the physical scales and interactions between the accreting plasma components. For instance, a rapid optical/infrared lag of $\sim$\,100\,ms relative to X-rays has now been observed in several systems and appears to be an important constraint for models of jet launching and acceleration \citep{gandhi_elevation_2017}.

    
    Yet, these studies rely on the source being both sufficiently bright and well-observed at multiple wavelengths simultaneously, the former being rare and the latter being marred by the inherent unpredictability of these outbursts. As such, these studies have so far been few and far between, and rarely carried out multiple times over the same outburst - though there are hints at an evolution of processes at different stages of the outburst \citep[see, e.g., ][though note that the latter compares two different outbursts]{veledina_swiftj1753ccfs_2017, Vincentelli_SubSecondVariability_2019}. Solutions are not yet unique, with processes such as a jet and a hot flow invoked to explain certain signatures on intermediate timescales \cite[e.g. ][]{veledina_accretion_2013, Malzac_GX339_2018}. We still remain severely data-limited in terms of high-quality strictly simultaneous multiwavelength time series in order to make progress.
    
    In 2018, one particular X-ray binary was discovered. It became bright enough and observed well enough that a good picture of its initial, several-week-long hard state -- including evolving inter-band correlations and Fourier components, observed at over 100\,Hz -- has been made possible.
    
	
	Discovered first as optical transient ASASSN-18ey on 6 March 2018 \citep{ATel11400, Tucker_ASASSN18ey_2018} and then as an X-ray source on 11 March \citep{ATel11399}, MAXI J1820+070 (hereafter J1820) was detected during the rapid outburst rise. It quickly rose to a brightness of $\sim$4 Crab \citep{Shidatsu_Monitoring_2019}, becoming the brightest extra-solar object in the X-ray sky by the time it peaked on 23 March \citep{corral-santana_blackcat:_2016}. By this point, its brightness had led to observations at many sites (e.g. \citealt{ATel11418, ATel11420, ATel11424, ATel11427}); not only did these observations quickly identify it as a likely Black Hole LMXB \citep{ATel11418, ATel11420, ATel11423}, but they also revealed rapid optical flaring (\citealt{ATel11421, ATel11426}) and even a significant optical/X-ray correlation \citep{Paice_J1820_2018}. Later, a QPO would be first identified in this source at around this peak \citep{ATel11488, ATel11510}, and would be seen to evolve over the next few months \citep{Stiele_Kong_J1820_Evolution_2020}. This stage of the outburst was the `hard state', where it is believed that the inner edge of the accretion disc is recessed, and a relativistic jet is present \citep{Done_EverythingAccretion_2007}.
	
\begin{figure}
		\includegraphics[width=\columnwidth]{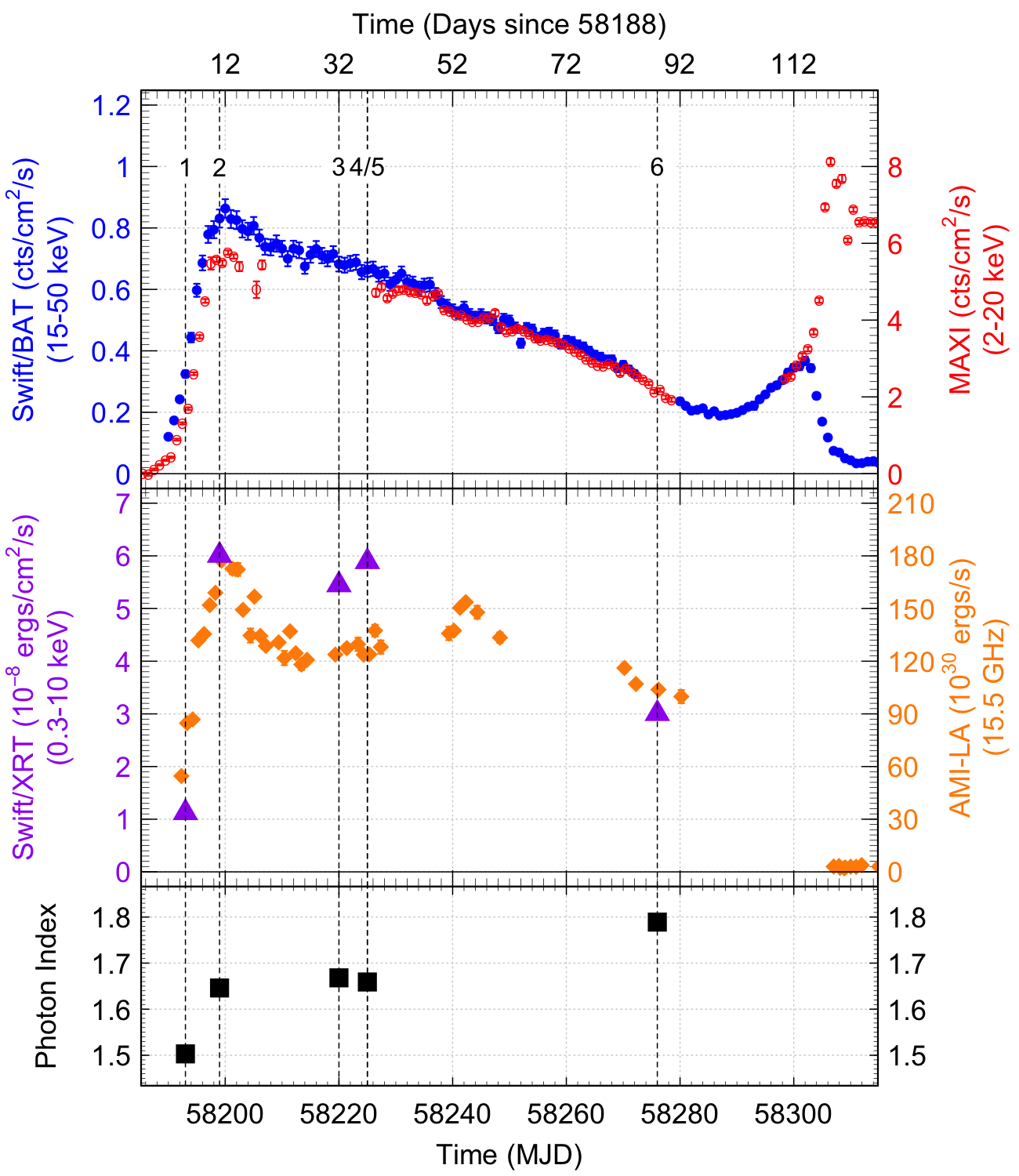}
		\caption{Timeline of MAXI\,J1820+070's 2018 outburst. Vertical black dashed lines denote dates of observations (see Table \ref{tab:observations}). The source transitions to the soft state just after MJD 58300. \textit{Swift}/XRT fluxes and photon indices were obtained from spectra produced by \textit{Swift}'s `Build XRT Products' tool \citep{Evans_SwiftXRTSpectra_2009}, using only data that corresponds to our epochs; errors are smaller than the marker size for both datasets. \textbf{Top:} Continuous monitoring done by \textit{Swift}/BAT (filled blue circles) and ISS/MAXI (open red circles). \textbf{Middle:} X-ray fluxes from \textit{Swift}/XRT (purple triangles). Also plotted are Radio luminosities from AMI-LA (orange diamonds), presented in \citet{Bright_J1820_2020}. \textbf{Bottom:} Photon indexes (black squares).}
		\label{fig:timeline}
	\end{figure}

	After the peak, J1820 entered a gradual decline in X-ray flux. In early July 2018, it transitioned rapidly to the soft state \citep{Homan_J1820_2018}, where the accretion disc extends to the Innermost Stable Circular Orbit (ISCO) and the jet is quenched. During this time, a unique blackbody X-ray emission signature was detected, which has been suggested as originating from within the ISCO, the so-called 'plunge region' \citep{Fabian_PlungeEmission_2020}. J1820 remained in this state until late September 2018, when it transitioned back to the hard state \citep{ATel12057}. It has since undergone a series of small `rebrightenings' \citep{ATel12567, ATel12732, ATel12747, ATel13014, ATel13025, ATel13502}, but as of yet, it has not undergone a second outburst. Fig. \ref{fig:timeline} shows a timeline of the hard state outburst at X-ray and radio wavelengths, using data from the Neil Gehrels \textit{Swift} Observatory, MAXI, and AMI-LA.
	
    
    Radio parallax measurements have since constrained J1820 to a distance of $2.96\pm0.33$\,kpc \citep{Atri_J1820Parallax_2020}, and the optical parallax found using Gaia EDR3 \citep{gaia_EDR3_2020} gives a distance of $2.94^{+0.87}_{-0.55}$\,kpc (calculated using the recommended zero-point correction -- \citealt{Lindegren_ZeroPoints_2020}), which improves on the previous estimate reported in Gaia DR2 \citep{Gandhi_GaiaDR2_2019}.  
	
	J1820's brightness led to several multiwavelength campaigns using high-time-resolution instrumentation over the course of its outburst. In \citet{Paice_1820Letter_2019}, we discussed the optical/X-ray correlations taken from a single night, using HiPERCAM and NICER during the rising hard accretion state. Therein, we noted the presence of a sub-second optical lag of order $\sim$\,100\,ms between the bands dependent upon wavelength, which we attributed to structure within the compact jet, and presence of an anti-correlation, which we put in the context of the hot accretion flow. Together with GX\,339--4 and V404\,Cyg \citep{gandhi_elevation_2017}, these results make J1820 the third well-studied black hole XRB to show the aforementioned sub-second lag. 
	
	The above results all highlight the importance of J1820 as a benchmark for understanding accretion. Here, we expand on these results to trace the timing properties through the primary hard state, including four new observations between NICER and another optical instrument, ULTRACAM, as well as a second correlated HiPERCAM/NICER observation later in the outburst. All observations were taken during the initial hard state, cover time resolutions from 8\,Hz to 300\,Hz, and cover a span of roughly 80 days in total. We construct a picture of the evolving optical/X-ray correlations over this period, and discuss to what processes they may relate.

	\section{Observations}
	
	\begin{table*}
    	\centering
	    \caption{Log of Simultaneous ULTRACAM/HiPERCAM \& NICER Observations for MAXI J1820+070 2018 Hard State Outburst.}
        \begin{tabular}{c c c c c c c c r r }
        \hline 
        Epoch & UT date & \thead{Day \\ Num.$^{1}$} & \thead{Time Start \\ (MJD-58000)$^{2}$} & \thead{Time End \\ (MJD-58000)$^{2}$} & \thead{Optical \\ Instrument} & \thead{X-ray \\ Instrument}    & \thead{Optical \\ Filters} & \thead{Correlated \\ Time (s)}  &  \thead{Cadence \\ (ms)$^{3}$} \\
        1 &  2018-03-16  & 5 & 193.38035 & 193.38518 & ULTRACAM & NICER & $u_sg_sr_s$ & 417 &  13.8 (138)\\
        2 &  2018-03-22  & 11 & 199.34590 & 199.41903 & ULTRACAM & NICER & $u_sg_si_s$ & 1279 &  9.06 (27.2)\\
        3 &  2018-04-12  & 32 & 220.40479 & 220.40736 & ULTRACAM & NICER & $u_sg_si_s$ & 222 &  10.4 (31.1)\\
        4 &  2018-04-17  & 37 & 225.17310 & 225.25110 & HiPERCAM & NICER & $u_sg_sr_si_sz_s$ & 1648 &  2.93\\
        5 &  2018-04-17  & 37 & 225.30454 & 225.31547 & ULTRACAM & NICER & $u_sg_si_s$ & 780 &  23.0 (69.0)\\
        6 &  2018-06-07  & 88 & 276.19520 & 276.22171 &  HiPERCAM & NICER & $u_sg_sr_si_sz_s$ & 1196 &  2.93\\
        \hline
        \end{tabular}
        \footnotesize{\textit{1} \quad Number of days since 2018 March 11 (MJD 58188), as used by \citet{Stiele_Kong_J1820_Evolution_2020}}\\
        \footnotesize{\textit{2} \quad Start and end of simultaneous times only -- observations may contain gaps.} \\
        \footnotesize{\textit{3} \quad Limiting cadences in optical. Numbers in brackets are $u_s$ band cadences, if different from the other bands.}

		\label{tab:observations}
    \end{table*}

	\subsection{NTT/ULTRACAM -- Optical}
	\label{sec:ULTRACAM}



    ULTRACAM is a fast-timing optical camera on the 3.58\,m New Technology Telescope (NTT) in La Silla, Chile. It was built for the purpose of fast optical timing in multiple wavebands. To this end, it includes three channels for simultaneous multiwavelength monitoring (with replaceable filters). It can also observe at frame-rates well above 100\,Hz; this is achieved by the lack of a physical shutter, and frame-transfer CCDs that can rapidly shift charge into a storage area for reading out, freeing up the original pixels for observation and thereby achieving low dead times \citep{dhillon_ultracam:_2007}.
	
	We used ULTRACAM to observe J1820's initial outburst peak in the early mornings of 2018 March 16, March 22, April 12, and April 17. All observations were carried out with the $u_s$, $g_s$, and $i_s$ SDSS filters, except for the first, which used the $r_s$ filter instead of $i_s$. Unlike most observations of this type, the times were not explicitly chosen to coincide with X-ray observations -- instead, the overlaps were purely coincidental and the serendipitous result of near-constant monitoring of J1820 by multiple telescopes. ULTRACAM was used in two-window mode (one each for the target and comparison star), with both window sizes of 50\,x\,50 pixels with a 2\,x\,2 binning for sensitivity and speed. See Table \ref{tab:observations} for observational details. J1820 was very faint in $u_s$, and so ULTRACAM's on-chip co-adding feature was used; this provides a longer exposure time in $u_s$ so as  to increase signal-to-noise ratio.
	
	
	The data were reduced using the ULTRACAM pipeline v9.14 \citep{dhillon_ultracam:_2007}. The bias was subtracted from each frame, and flat field corrections were also applied. Aperture sizes scaled to the instantaneous seeing were used, with radii between 0$\farcs$7 and 3$\farcs$5, with an annulus of between 12$\arcsec$ and 6$\farcs$3 to calculate the background. These apertures had variable centre positions that tracked the centroids of the sources on each frame, with a two-pass iteration (where an initial pass is made to track the sources on the CCD before a second photometry pass) used for accuracy. Our times were then adjusted to Barycentric Dynamical Time (BJD\_TDB) using methods given in \cite{eastman_achieving_2010}. 
	
	Our comparison star is located at RA = 18 20 26.43, Dec = 07 10 11.7 (J2000), and is listed in the \textsc{PanSTARRS} survey catalog \citep{Magnier_PanSTARRS_2020} with $g_s$/$i_s$ magnitudes of 13.3083/12.233 respectively. The star was taken to be constant, and was used for photometric calibration. For the $u_s$ observations, the comparison star was too faint to perform photometry within a single frame. Hence, we used the measured zero-point magnitude for the $u_s$ band in photometric conditions for ULTRACAM (Vik Dhillon, priv. communication) in order to calibrate our observations. We extracted the J1820 and comparison star magnitudes using aperture photometry with a variable aperture size that was dictated by the seeing conditions. The aperture also tracked the centroid of the source of interest by using a bright star in the field as a reference. For the $u_s$ observations, we used J1820 as the reference object itself so as to not lose tracking within the field.
	
    

	\subsection{GTC/HiPERCAM -- Optical}
	\label{sec:HiPERCAM}
	
    High-speed multi-colour photometry of J1820 was carried out using HiPERCAM \citep{Dhillon_First_2018} on the 10.4\,m Gran Telescopio Canarias on La Palma. HiPERCAM uses 4 dichroic beamsplitters to image simultaneously 5 optical channels covering the $u_sg_sr_si_sz_s$-bands (respectively, central wavelengths 3526, 4732, 6199, 7711 and 9156\,\angstrom). The CCDs were binned by a factor of 8 and used in the highest-speed drift mode. We orientated the instrument (PA = 58$^{\circ}$) and used two windows (96x72 pixels each), one centered on J1820, and another on a comparison star, APASS--34569459 \citep{Henden_APASS_2015}. The observations discussed here were taken on 2018 April 17 from 03:26--06:11 UT, and 2018 June 07 from 04:41--05:39, coordinated with NICER. The exposure time was 2\,ms, the cadence 2.9\,ms, the median seeing 2.2\arcsec. The sky was affected by mild cirrus on both dates.
    
	We used the HiPERCAM pipeline software\footnote{\url{https://github.com/HiPERCAM/hipercam}} to de-bias, flat-field and extract the target count rates using aperture photometry with a seeing-dependent circular aperture tracking the centroid of the source. Sky background was removed using the clipped mean of an annular region around the target. The target was brighter than all stars in the field. We thus used the raw target counts for the analyses presented herein; note that our primary results are not affected when using photometry relative to the comparison star.

	\subsection{ISS/NICER -- X-ray}
	\label{sec:NICER}
	
	NICER (Neutron star Interior Composition ExploreR) is an X-ray instrument aboard the International Space Station (ISS). It comprises 52 functioning X-ray concentrator optics and silicon drift detector pairs, arranged in seven groups of eight. Individual photons between 0.2-12 keV, and their energies, can be detected to a time resolution of 40\,ns \citep{GendreauArzoumanian_NICER_2016}.
	
	J1820 was observed with an intensive monitoring program during the initial hard state of its outburst. Data reduction of ObsIDs 1200120105, 1200120107, 1200120127, 1200120131, and 1200120172, were completed using {\sc nicerdas}, a collection of NICER-specific tools, and part of HEASARC\footnote{\url{https://heasarc.gsfc.nasa.gov}}. Full Level2 calibration and screening was conducted with {\textit {nicerl2}}, which calibrated, checked the time intervals, merged, and cleaned the data. Barycentric correction was carried out using {\sc barycorr}, then the photon events (all between 0.2-12\,keV) were binned to the times of the optical light-curve.
	
	
	
	\section{Method} \label{sec:method}
	
	Our analysis of the optical and X-ray data involves creating simultaneous lightcurves, Cross-Correlation Functions (CCFs) and Fourier analysis. In the following we detail the methodology used.
	
	\subsection{Simultaneous Lightcurves} \label{sec:simultaneous_lightcurves}
	
	Simultaneous lightcurves are plotted in Fig. \ref{fig:Lightcurves}. The optical and X-ray data are not, by default, binned simultaneously. However, while the optical data were taken in discrete time bins by both instruments, NICER is a photon-counting instrument and thus records the arrival time of each photon.
	Therefore, we create simultaneous lightcurves by binning the photons directly to the optical time bins, after barycentering both datasets. Since the optical lightcurves have a constant deadtime (time between the bins in which no data were recorded) the X-ray photons observed during this time are disregarded. For X-rays, the square root of the counts per bin was used to determine the error for each bin. Since the $u_s$ band data were sampled at a different rate to the other optical bands, a separate X-ray lightcurve was created. This lightcurve is not plotted in Fig. \ref{fig:Lightcurves}, but was used in creating the Cross-Correlation functions and in the Fourier analysis for the $u_s$ band data in epochs 1--3 and 5.

	\subsection{Cross-Correlation Functions}
	
    Cross-correlations are plotted in Fig. \ref{fig:10s_CCFs} \& \ref{fig:2s_CCFs}. Cross-correlations are measurements of how much one lightcurve (or any time-series) varies dependent on another as a function of lag. In these cases, we create optical vs. X-ray cross-correlations; the figures therefore show the response of the optical lightcurves to variations in the X-ray lightcurve, as a function of time lag. Positive values indicate a net correlation at that lag, and negative values a net anti-correlation, each normalised so that 1 and -1 indicate perfect correlations and anti-correlations.
    
    The cross-correlations were produced by splitting the simultaneous lightcurves into segments of equal length. Each segment was then `pre-whitened' by removing a linear trend. A Cross-Correlation Function (CCF) was then run on each segment, using the methodology of \citet[P. 390]{VenablesRipley_ModernAppliedStatistics_2002}. The mean CCF was then determined and the standard error on each bin was calculated. To probe variations on different timescales we compute CCFs using segment sizes of 10\,s (Fig. \ref{fig:10s_CCFs}) and 2\,s (Fig \ref{fig:2s_CCFs}).
	
	\subsection{Fourier Analysis}
	
	Fourier analysis is presented in Fig. \ref{fig:powerspectra}--\ref{fig:time_lags}. These involved computing the Fourier transform of the lightcurves and then analysing them at each frequency.
	
	The power spectra represent the amplitude of the variability at each Fourier frequency. The coherence represents the relative magnitude of the complex-valued cross-spectrum, i.e. a measure of how the bands are correlated at that frequency. The phase lags represent the relative phase angle of the complex-valued cross-spectrum, i.e. a measure of the lag between the bands at each frequency as a function of phase (measured in radians). The time lags show the same data as the phase lags, but converted into the time domain.
	
	This analysis made use of the Stingray\footnote{\url{https://github.com/StingraySoftware/stingray}} python package \citep{Huppenkothen_Stingray_2019}. Values for the intrinsic coherence, and errors on those values, were determined using methods described in \cite[Eqn. 8]{Vaughan_Nowak_1997}, where our data fit into the category of `High powers, high measured coherence'.
	
	Good Time Intervals (GTIs) were used based on the individual epochs of X-ray observation, and then cross-spectra were computed over independent lightcurve segments and averaged. The segment lengths were $2^{12}$ bins for epochs 1--3 and 5, and $2^{14}$ bins for epochs 4 and 6. For observations with co-adding in $u_s$, the nearest multiple of 2 was used as the bin length, so that the lightcurve segments were of similar size compared to the other filters of the same observation. These segment sizes were selected to balance frequency range against statistics, making sure that all bands were averaged over at least 5 segments (aside from the $u_s$ bands in epoch 3 and 5, which had only 3 and 4 segments respectively).
	
	Root-mean-squared (rms$^2$) normalisation was applied to the power spectra \citep{Belloni_Hasinger_Aperiodic_1990}. The white noise was fitted and removed from the power spectra before calculating the coherence (see Section \ref{sec:powerspectra} for details). In Figures 5--8 the frequency-dependent products were binned logarithmically in frequency; for the power spectra the factor was 1.1, while for the coherence, time lags, and phase lags the factor was 1.3 (these were chosen to balance the clarity of features with the size of the uncertainties).
	
	Time lags were calculated by dividing the phase lags by $2 \pi f$, where $f$ is the frequency of the bin. Since the conversion is ambiguous and could be $\pm 2\pi$, we assumed that the phase lags of the frequency bins around 1\,Hz were correct, based on their relationship to the sub-second time lag seen in Figure \ref{fig:2s_CCFs}. Each time lag was then arbitrarily shifted based on what would cause the fewest discontinuities.

	\section{Results} \label{sec:results}
	
	In the Figs. \ref{fig:Lightcurves}--\ref{fig:time_lags}, the violet plot on the left shows the timeline of the outburst in MJD, seen by \textit{Swift}/BAT (see Figure \ref{fig:timeline}) -- the stronger the colour, the brighter J1820 was in hard X-rays. The epochs are marked. Each plot shows the variation in all bands. The colour key is as follows: $u_s$ (blue), $g_s$ (green/teal), $r_s$ (red), $i_s$ (dark red/brown), $z_s$ (black), and X-rays (violet).
	
	\subsection{Lightcurves} 

	\begin{figure*}
		\includegraphics[width=\textwidth]{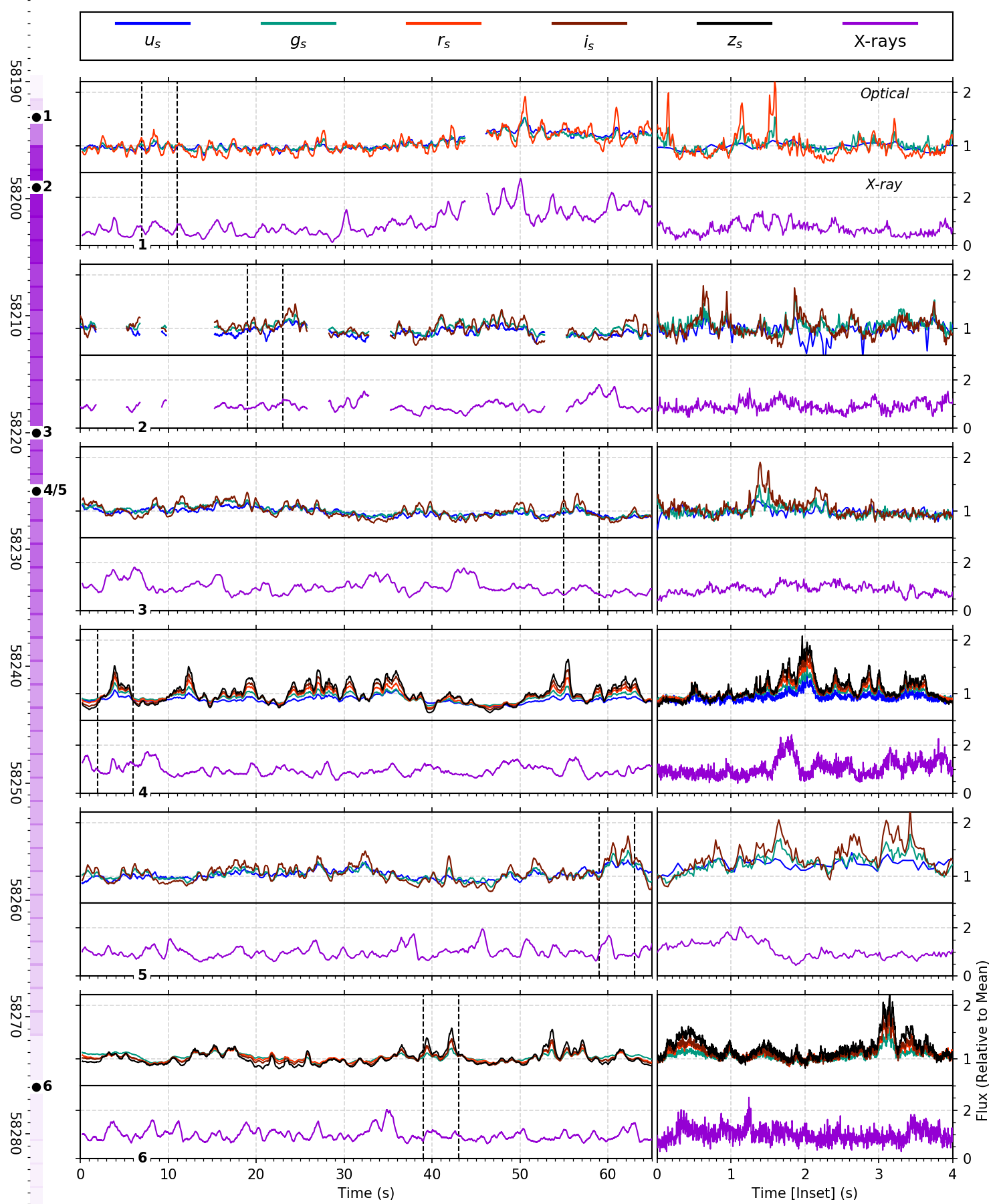}
		\caption{Portions of Optical and X-ray lightcurves from each epoch, normalised to each band's mean. \textbf{Far Left:} Timeline of the outburst in MJD, seen by \textit{Swift}/BAT (stronger colour = brighter in hard X-rays). \textbf{Left:} 60\,s overview, binned to show overall variations. \textbf{Right:} 4\,s insets, marked by dotted lines in the left. Arbitrary offsets have been applied to the time axis. The $u_s$ band for Epoch 6 suffered from poor statistics, and was thus not plotted.}
		\label{fig:Lightcurves}
	\end{figure*}
	
	Fig. \ref{fig:Lightcurves} shows portions of the lightcurves from each of the epochs in all optical bands as well as in X-rays. The lightcurves show a lot of similarities -- in the optical there are numerous sub-second flares with an increase of a factor $\sim$1.5--2 in flux. A common property of all lightcurves is that the variations tend to be far stronger in the red than in the blue, and is particularly true of the sub-second flares -- this is also seen in other hard-state LMXBs \citep{gandhi_rapid_2010, gandhi_furiously_2016}. These flares become less frequent as the epochs continue, but are still present in epoch 6. Interestingly, the lightcurves are sometimes anti-correlated during these flares, with optical activity rising while X-ray activity decreases - see, for example, the inset to Epoch 4.

	\subsection{Cross-Correlation Functions} \label{ccf}

	\begin{figure}
		\includegraphics[width=\columnwidth]{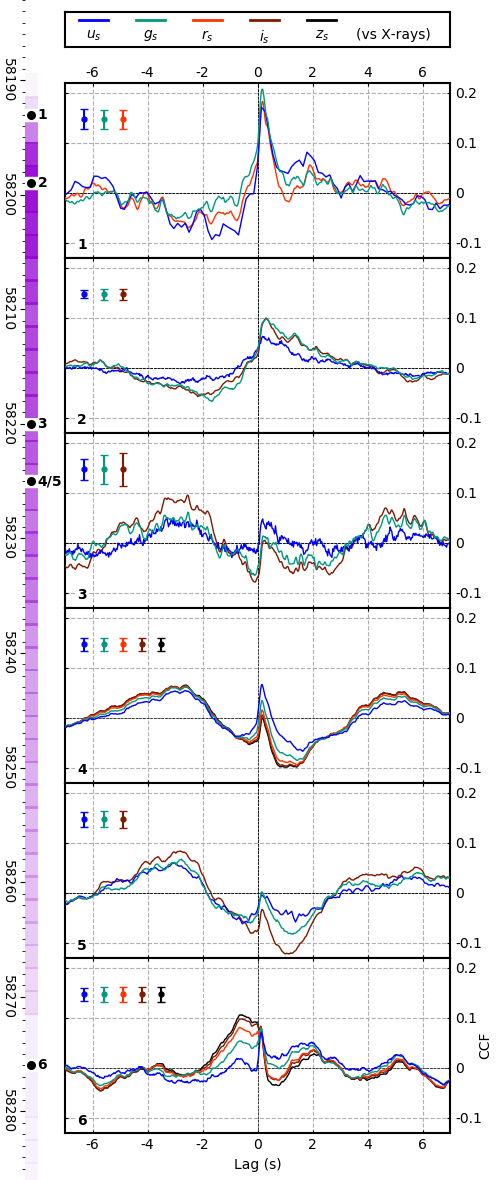}
		\caption{Optical/X-ray cross-correlation functions (i.e. a peak/trough at positive lags means that the optical flux lags the X-ray flux). The CCFs shown are the average CCF computed from multiple 10\,s segments. The lightcurves were binned to roughly match the lowest time resolution (epoch 5, 43\,Hz) in order to better compare CCF coefficient values (except for $u_s$ in epochs 1-3 and 5, due to their significantly lower time resolution). Standard errors, averaged over the plotted range, are shown. \textbf{Far left:} Timeline of the outburst -- see Fig. \ref{fig:Lightcurves} caption.}
		\label{fig:10s_CCFs}
	\end{figure}

	\begin{figure}
		\includegraphics[width=\columnwidth]{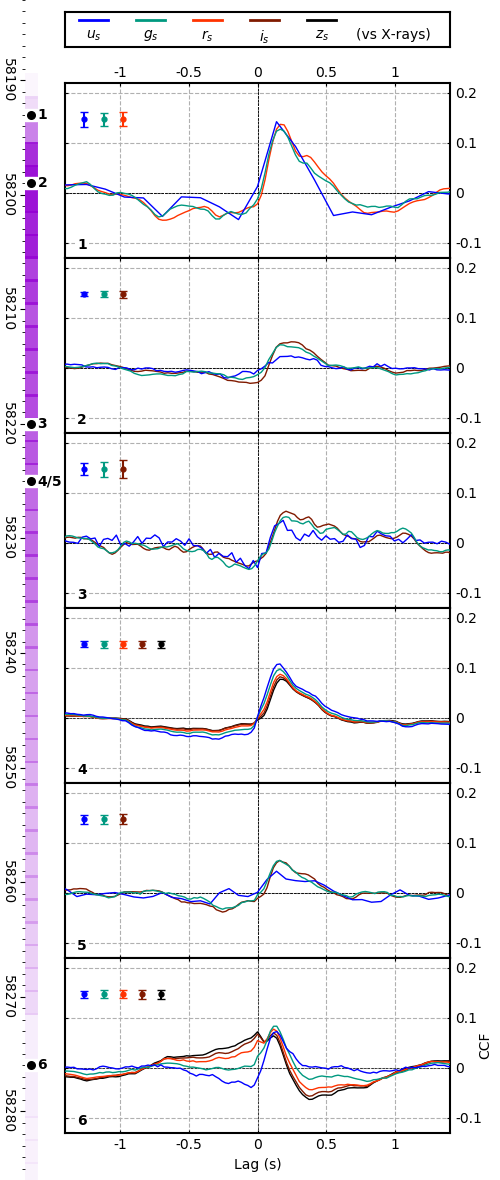}
		\caption{Same as in Fig. \ref{fig:10s_CCFs}, but the CCFs are created from 2\,s segments instead. This allows us to better compare the rapid variability, and shows that the sub-second peak at positive lags is present in a similar place in every single band and epoch.}
		\label{fig:2s_CCFs}
	\end{figure}
	
	CCFs from 10\,s segments can be seen in Figure \ref{fig:10s_CCFs}. Each epoch shows a sub-second correlation peak. Epochs 1--5 also show some form of a `precognition dip'; i.e. an anti-correlation at negative lags, which means that either the optical lightcurve dips a few seconds before an X-ray flare, or that optical flares occur before an X-ray dip.
	
	We can also see how the CCFs evolve over time. Between epochs 1--5, the correlation steadily decreases at positive lags, perhaps caused by an anti-correlation component becoming more significant in the data. Additionally, from epoch 3 onwards a new positively-correlated component appears, peaking at negative lags, which is stronger at longer wavelengths. This feature evolves from $\sim$-3\,s in epoch 3 to -1\,s in epoch 6.
	
	To probe the more rapid variations, we also created CCFs from 2\,s segments which can be seen in Figure \ref{fig:2s_CCFs}. These reveal several details. Firstly, the sub-second correlation is strongest just before the peak of the outburst. Secondly, the lag of the CCF peak is fairly constant from epoch 2 onwards, even into epoch 6. Finally, the CCF peak has a `tail' extending from the initial peak out to 0.5--0.75\,s and appears to shrink over time, or at least become less significant compared to some anti-correlated component; see, in particular, the difference between epochs 4 and 6.
	
	The sub-second correlation was previously reported in \citet{Paice_1820Letter_2019}. In that work, we found that the lag was greater at longer wavelengths -- this can be seen in several of these epochs, and will be shown more explicitly in Section \ref{sec:time_lags}, and finally discussed in Section \ref{sec:compact_jet}. Our highest-resolution epochs, 4 and 6, also show a small spike in the CCF at 0\,s lag. This is confirmed by CCFs created from 1\,s lightcurve segments and below. However, our data do not have sufficient time resolution to study these particular features.
	
	A closer inspection shows that the correlated negative-lag component is present in epoch 6, but with curious results. Firstly, the peak is now at 0\,s, not -1\,s. Secondly, at negative lags the longer wavelengths have a stronger correlation, while the shortest wavelength ($u_s$) shows a trend towards \textit{anti}-correlation. At positive lags, the reverse is true; it is the longer wavelengths that now show a trend towards anti-correlation. This shows either that this negative component affects the signals down to these rapid timescales, or that there are two components that affect these lags -- this correlated negative component, and a new component that has appeared between epochs 5 and 6. These possibilities will be discussed in Sections \ref{sec:Negative_Lag} and \ref{sec:epoch6}.
	
	\subsection{Fourier Analysis} \label{sec:fourier}
	
	To better understand the nature of different components contributing to the CCF, we perform the decomposition of the observed variability into different time scales using Fourier technique.
	Figs. \ref{fig:powerspectra} -- \ref{fig:time_lags} show various Fourier components in each optical band plotted against X-rays.
	
	
	
	
	\begin{table*}
    	\centering
	    \caption{Fitting parameters for the power spectra. Each cell contains the following: Number of Lorentzians fitted; white noise level in fractional rms$^2$ units $\times 10^{-5}$ (Reduced Chi$^2$ $\chi^2_\nu$).}
        \begin{tabular}{c c c c c c c c c c c c c c c c c c c }
        \hline 
        Epoch & X-rays & $u_s$ & $g_s$ & $r_s$ & $i_s$ & $z_s$ \\
        1 &  2; 36.6 (1.47) & 2; 59.4 (0.561) & 2; 6.19 (0.847) & 2; 4.02 (1.43) & &\\
        2 &  2; 36.9 (1.4) & 4; 84.2 (1.13) & 4; 6.95 (2.84) & & 4; 3.52 (1.69) &\\
        3 &  2; 11.7 (0.992) & 2; 26.0 (0.495) & 4; 5.79 (1.08) & & 4; 3.92 (0.894) &\\
        4 &  5; 15.4 (0.86) & 3; 0.77 (2.07) & 6; 0.09 (3.98) & 6; 0.11 (4.98) & 7; 0.14 (7.77) & 8; 0.23 (8.93)\\
        5 &  2; 11.5 (1.16) & 2; 32.8 (0.484) & 4; 5.8 (0.762) & & 4; 3.25 (1.34) &\\
        6 &  5; 29.6 (1.66) & 2; 324 (0.65) & 4; 0.38 (2.56) & 4; 0.57 (2.76) & 4; 0.66 (2.71) & 4; 10.1 (4.28)\\
        
        \hline
        \end{tabular}
		\label{tab:Lorentzians}
    \end{table*}

	\subsubsection{Power Spectra} \label{sec:powerspectra}
	
	\begin{figure}
		\includegraphics[width=\columnwidth]{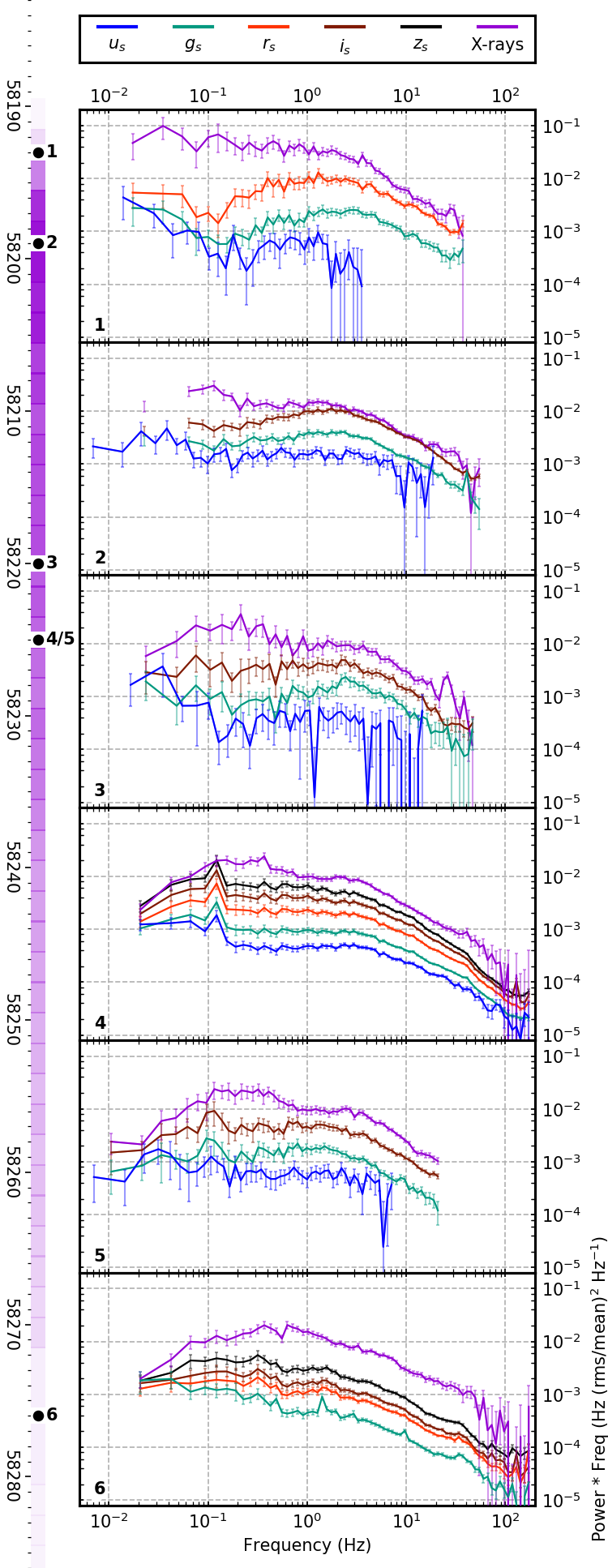}
		\caption{Power spectra of J1820. White noise has been fitted and removed from each. \textbf{Far left:} Timeline of the outburst -- see Fig. \ref{fig:Lightcurves} caption.}
		\label{fig:powerspectra}
	\end{figure}
	
	The power spectra can be seen in Fig.~\ref{fig:powerspectra}. In addition to the five optical bands (detailed at the beginning of Section \ref{sec:results}) the X-ray power spectra are also shown.\footnote{Note that these are only for the X-ray data that are strictly simultaneous with our fast optical photometry (aside from the $u_s$ band in cases of co-adding, i.e. epochs 1--3 and 5)}, and not the full spectra from the NICER observations. Thus, these are not directly comparable to the power spectra in \citet{Stiele_Kong_J1820_Evolution_2020}. The power is in fractional rms$^2$ units and is multiplied by the frequency.
	
	As noted in Section \ref{sec:simultaneous_lightcurves}, the separate X-ray lightcurves for the co-added $u_s$ bands in epochs 1--3 and 5 are not shown. Additionally, the $u_s$ power spectrum in epoch 6 is not shown due to the poorer data quality. 
	
    A mix of zero-centered and non-zero-centered Lorentzians along with a constant white noise component were fitted to each band. For these plots, that white noise component was removed and the fitted parameters can be seen in Table~\ref{tab:Lorentzians}. The increased numbers of Lorentzians (and increased $\chi^2_\nu$ values) for epochs 4 and 6 are due to the higher cadences, larger segment sizes, and lower noise levels in the HiPERCAM data; these lead to far lower uncertainties, and thus require more Lorentzians to fit numerous features in these bands.
    
	\textit{Regarding the evolution of the power:} In all epochs, the power in the optical bands is consistently higher at longer wavelengths, although highest overall in X-rays. The manifestation of this can be seen in Figure \ref{fig:Lightcurves}, where one can see activity at longer wavelengths being much stronger than that at shorter ones.
	
	The power in each band evolves over the course of the outburst. At optical wavelengths, the power above $\sim$3\,Hz drops between epochs 1 and 6 by almost an order of magnitude. This is most evident when looking at 10\,Hz in the $i_s$-band power spectrum. However, at the lowest frequencies it appears more stable. This could be interpreted as a Lorentzian component peaking at $\sim$1--2\,Hz and becoming less significant as the outburst continues. However, this does not mean that the component disappears. Furthermore, a small feature is seen to peak at $\sim$30--40\,Hz in all optical bands in epochs 4 and 6 (the only bands that extend to this frequency with good statistics. Epochs 2 and 3 may show this too, but the uncertainties are too large to confirm this).
	
	Meanwhile, the X-ray power spectrum behaves in the opposite manner. It remains roughly constant between epochs at all frequencies except the lowest, where it drops by an order of magnitude between the earliest and latest epochs.
	
	All the power spectra show a break at around 1\,Hz, and epochs 4 and 6 possibly show higher-frequency breaks at around 40\,Hz. However, Lorentzian fitting could not sufficiently quantify these breaks, and therefore their validity and cause will instead be left as a topic for future work.
	
	
	
	
	\textit{Regarding the existence of a Quasi-Periodic Oscillation:} In epochs 4 and 5, a Quasi-Periodic Oscillation (QPO)-like feature can be seen at $\sim$0.1\,Hz. While Lorentzian fitting did not significantly improve with an additional component at these frequencies for all bands, an X-ray QPO at these frequencies has been previously detected; the existence and effects of such a feature are discussed in Section \ref{sec:Negative_Lag}.
	
	\subsubsection{Coherence} 
	
	\begin{figure}
		\includegraphics[width=\columnwidth]{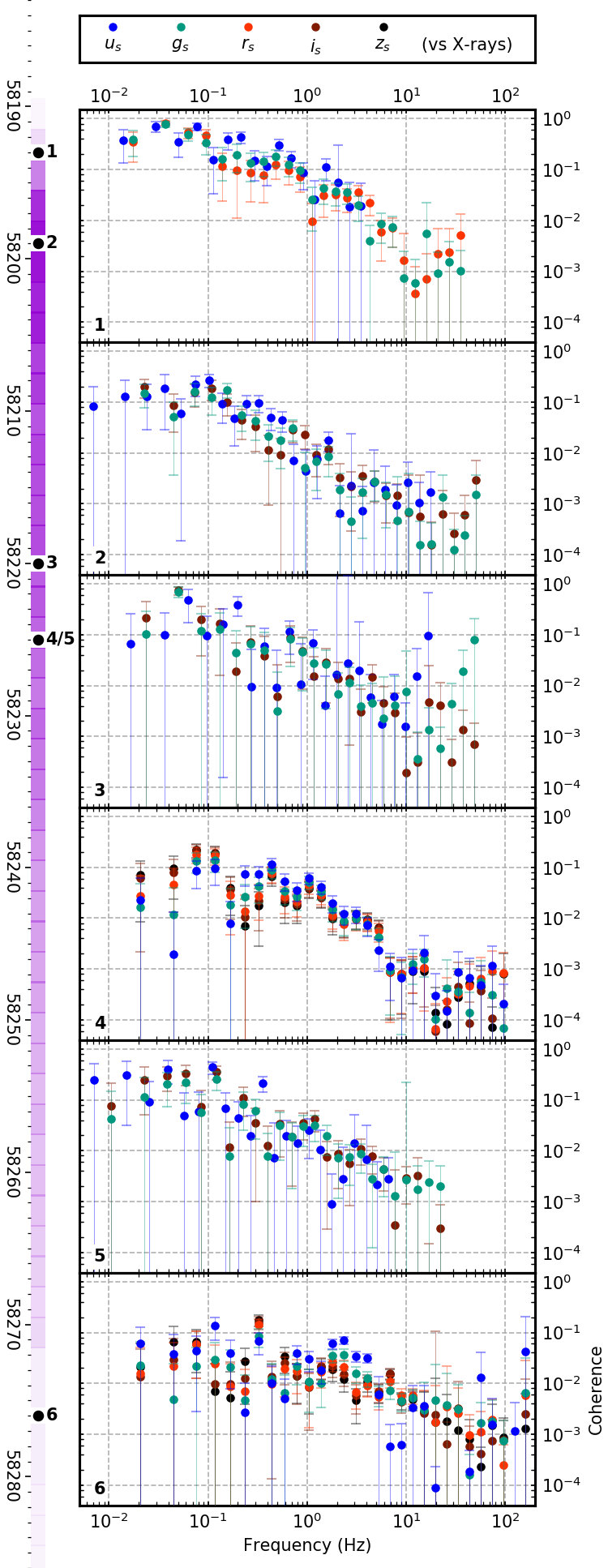}
		\caption{Coherence of J1820 over frequency, with the same rebinning as in Figures \ref{fig:powerspectra} \& \ref{fig:time_lags}.}
		\label{fig:coherence}
	\end{figure}

	Fig.~\ref{fig:coherence} shows the coherence. Overall, this is generally low (<0.1) at all frequencies, which is typical for these sources \citep[See, e.g., ][]{Malzac_GX339_2018}. During all epochs and bands, the optical is more coherent with the X-rays at lower frequencies, and decreases with increasing frequency in every epoch. However, the coherence at lower frequencies decreases as the outburst continues, eventually dropping by over an order of magnitude by epoch 6.
	
	There are numerous smaller features here, but for this work, we will just note the peaks which occur in the later epochs -- at 0.1\,Hz in epochs 4 and 5, and at 0.3\,Hz in epoch 6. These will be referred to later in Section \ref{sec:Negative_Lag} in the context of a QPO. 
	

    While there is no one relation for the dependence of coherence with optical band, there are discrete sections that do show clear trends. Saliently, in the 1--5\,Hz range, shorter wavelengths tend to be more coherent than longer ones (particularly in the epochs with the best statistics, such as 4 \& 6) -- this will be discussed in Section \ref{sec:time_lags}. There are also sections where the opposite is true -- spikes in coherence at the QPO frequency in epochs 4 and 6 are stronger at longer wavelengths. These, again, will be noted in Section \ref{sec:Negative_Lag}.

	\subsubsection{Phase Lags} \label{sec:phase_lags}
	
	\begin{figure}
		\includegraphics[width=\columnwidth]{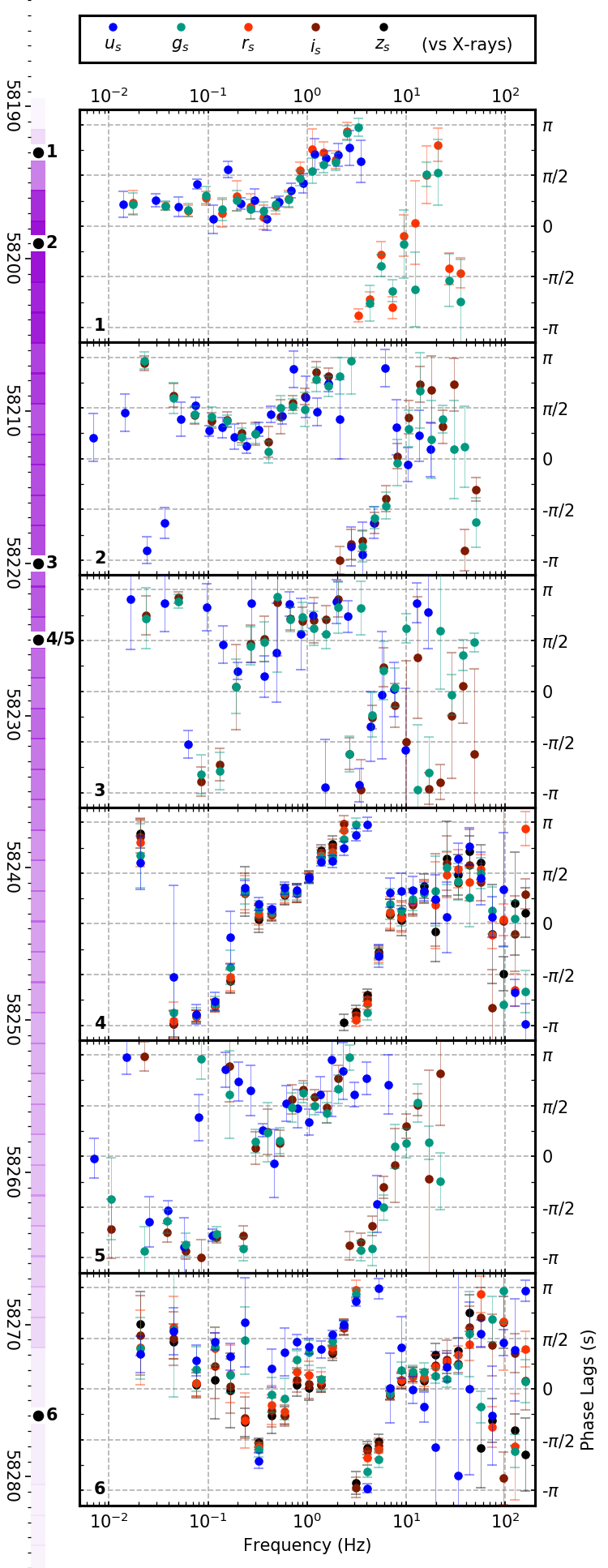}
		\caption{Phase lags of J1820 over frequency, with the same rebinning as in Figures \ref{fig:powerspectra} \& \ref{fig:time_lags}.}
		\label{fig:phase_lags}
	\end{figure}
	
	The phase lags can be seen in Figure \ref{fig:phase_lags}. Those in the range 1--10\,Hz are roughly the same across all observations, with a shift of +$\pi$ appearing at around 3\,Hz; these reflect the presence and stability of the positively-correlated peak. Above 10\,Hz, there are few clear trends and it is difficult to make definitive claims; if this regime is dominated by components with <0.1\,s delay, then we have many jumps from +$\pi$ to -$\pi$ over this period, and log binning would average out this behaviour.
	
	However, one difference is the behaviour of the phase lags below 0.5\,Hz. In epoch 1, the phase lags are mostly constant at +$\pi$/4, and in epoch 2, they appear to increase towards lower frequencies. However, in epochs 3-5 (a month after outburst peak), phase lags change to roughly $\pm\pi$ -- i.e. the two components are roughly in `anti-phase', where the peak of one component coincides with the trough of another (this is the Fourier representation of the anti-correlation component that appears in the CCF -- see Figure \ref{fig:10s_CCFs}). The transition to this anti-correlation in the phase lags occurs at around 0.2\,Hz, where there is a sudden discontinuity; analyses of epochs 4 and 5 are inconclusive in showing whether phase lags increase from $-\pi$, or decrease from $+\pi$ at this discontinuity. It is perhaps worth noting that negative phase lags, sometimes approaching $\pm\pi$, are seen at lower frequencies in multiple other sources  \citep[see ][]{gandhi_rapid_2010, veledina_swiftj1753ccfs_2017, Malzac_GX339_2018, Vincentelli_J1535_2021}.
	
	This lower-frequency behaviour then changes again much later in the outburst during epoch 6, at which time the anti-correlation component is now bounded to a small section at roughly 0.3\,Hz, with lower frequencies being generally above 0.

	\subsubsection{Time Lags} \label{sec:time_lags}
	
	\begin{figure}
		\includegraphics[width=\columnwidth]{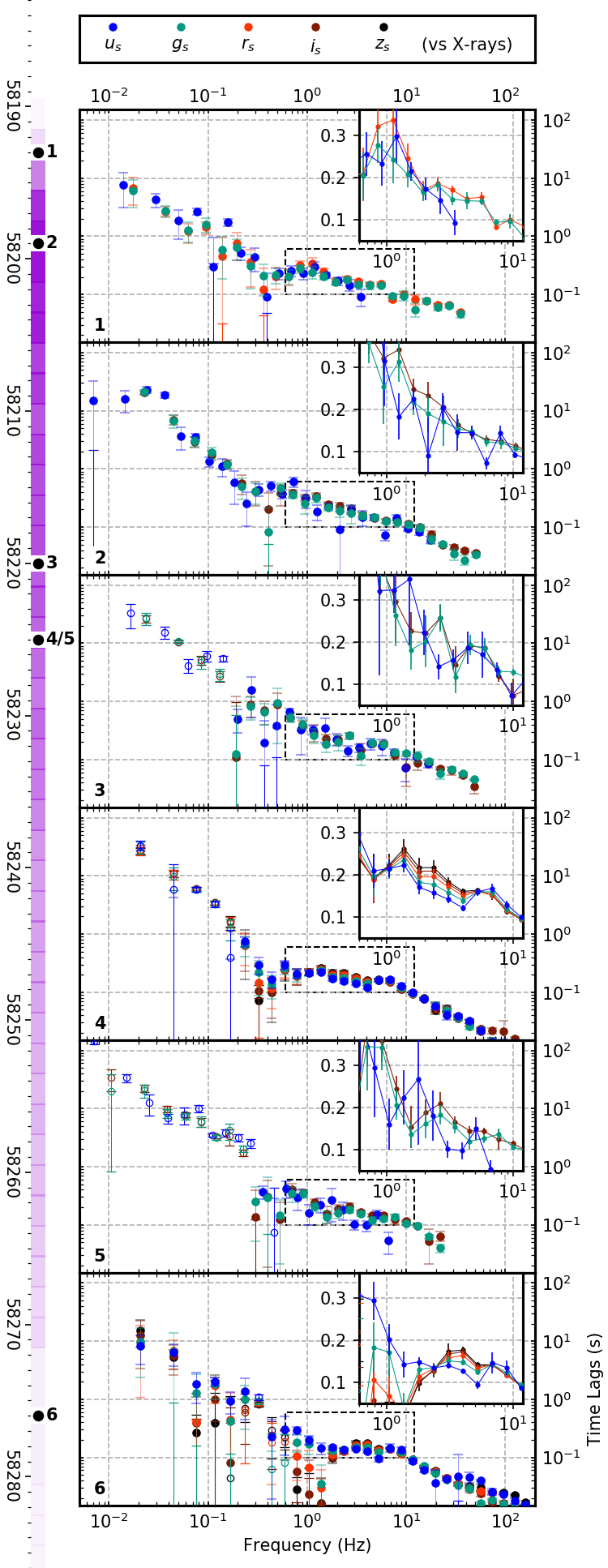}
		\caption{Time lags of J1820 over frequency. Inset for each plot shows a linear version of the region around the 0.2\,s lag seen in the CCFs. Open circles denote negative optical time lags.}
		\label{fig:time_lags}
	\end{figure}
	
    The time lags can be seen in Figure \ref{fig:time_lags}. At frequencies below $\sim$0.2\,Hz in epochs 3--5, there is confusion as to whether the time lags are positive or negative -- this depends on whether the phase lags are assumed to be positive or negative, which is unclear from Fig. \ref{fig:phase_lags}, as this is the point at which the phase lags are close to $\pm\pi$. 
	
	Figure \ref{fig:time_lags} also presents insets over the 1--10\,Hz range, showing the similarities over the epochs. Shorter frequencies almost uniformly have a smaller lag than longer frequencies over this range; this is only not the case in epochs with poorer statistics (i.e. epoch 3) or below 2.5\,Hz in epoch 6. This wavelength dependence will be discussed in Section \ref{sec:compact_jet}, with epoch 6 in particular discussed in Section \ref{sec:epoch6}.

	\subsection{Quasi-Periodic Oscillation -- Its Evolution and Lags} \label{sec:Negative_Lag}
	
	In epochs 4--6, a feature can be seen that is similar to a QPO, with significant effects in the coherence and the lags. In epochs 4 and 5, this feature is at roughly 0.1\,Hz, which increases to 0.3\,Hz in epoch 6. Each bin with this feature shows (i) an increase in the power spectra, (ii) higher overall coherence (sometimes by an order of magnitude, particularly in epochs 4 and 6), (iii) greater coherence at longer wavelengths, (iv) small error bars in the lags, and (v) negative time lags (changing from -4\,s in epoch 4 to -1\,s in epoch 6). These features are best seen in epochs 4 and 6, where the statistics are better than other epochs.
	
	This possible QPO can also be seen in the CCFs (Figure \ref{fig:10s_CCFs}). A positively correlated component can be seen between -4 and -3\,s in epochs 3--5, and at -1\,s in epoch 6, as indicated by the time lags seen in Fig. \ref{fig:time_lags}, often stronger at longer wavelengths. We briefly analysed the CCFs to test for the significance of this feature -- see Section \ref{sec:CCF_Sims} in the Appendix.
	
	
	
	As it turns out, a feature at this frequency is not a new discovery; \citet{Stiele_Kong_J1820_Evolution_2020} showed the evolution of a QPO in X-rays over time that corresponds exactly with our feature described here. Therefore, there appears to be a connection with this QPO and the features, including a negative lag in the CCF, in our data. Indeed, QPOs have been associated with changes in the lags in other LMXB sources previously \citep{Veledina_J1753QPO_2015, Malzac_GX339_2018, Vincentelli_J1535_2021}.
	
	Does this mean that the QPO shows optical variability preceding X-ray? Not necessarily; due to the periodic nature of phases (as discussed in Section \ref{sec:phase_lags}), phase lags between $\pi$--$2\pi$ radians would be represented as negative lags between -$\pi$--0, and this might be occurring here. Additionally, the negative lags seen in the CCF could just be a result of the periodic nature of this component; epochs 4 and 6, for example, show a second feature at positive lags (5\,s and 2\,s respectively). These give a time period of 8\,s and 3\,s respectively between the two features; this matches the period of the QPO in both epochs (roughly 0.125\,Hz and 0.3\,Hz respectively)\footnote{We also see this behaviour in epoch 3, where we do not see clear similar QPO features. However, the QPO is still detected by \citet{Stiele_Kong_J1820_Evolution_2020} during this time at a similar frequency. Additionally, with only 222s of correlated time, epoch 3 has the poorest statistics of any of our epochs; this may explain why we do not see such QPO features.}. See also Section \ref{sec:epoch6}, where we simulate the Fourier components of epoch 6, and show how both the positive and negative correlations disappear when the QPO's Fourier components are removed. There is no clear mechanism by which a QPO would directly cause X-ray emission to lag optical emission in this way, while there are a number of models that would show the opposite (e.g. see Section \ref{sec:hot_flow}); we thus consider the latter case to be the more likely one here.

	\subsection{The Oddity of Epoch 6} \label{sec:epoch6}
	
	The QPO described in the previous section cannot, by itself, explain all the lags in epoch 6; the phase lags that are significantly different from other epochs extend over the frequency range 0.08--2.5\,Hz, not just around the QPO frequency. At these frequencies, shorter wavelengths have a consistently \textit{greater} lag than longer wavelengths; this is the inverse trait of the sub-second lag seen between 1--5\,Hz in other epochs (while this sub-second lag and wavelength dependence is still seen in epoch 6, note also how this new component supersedes it up to 2.5\,Hz -- Fig. \ref{fig:time_lags}).
	
	The epoch 6 lightcurves show low coherence compared to other epochs ($\sim$ 0.01 -- the only exception here is the 0.3\,Hz frequency bin coincident with the QPO). As for the lags, over this range, $g_{s}$, $r_{s}$, $i_{s}$ and $z_{s}$ bands even have \textit{negative} lags with respect to X-rays, whereas $u_{s}$ almost always has positive lags at the same frequency. This behaviour is also evident in the 2\,s CCFs (Figure \ref{fig:2s_CCFs}), where the longer-wavelength $r_{s}$, $i_{s}$ and $z_{s}$ bands show a rising correlation at negative lags and peak at 0\,s, while the shorter-wavelength $g_{s}$ band does not, and the $u_{s}$ band shows an anti-correlation.
	
\begin{figure}
		\includegraphics[width=\columnwidth]{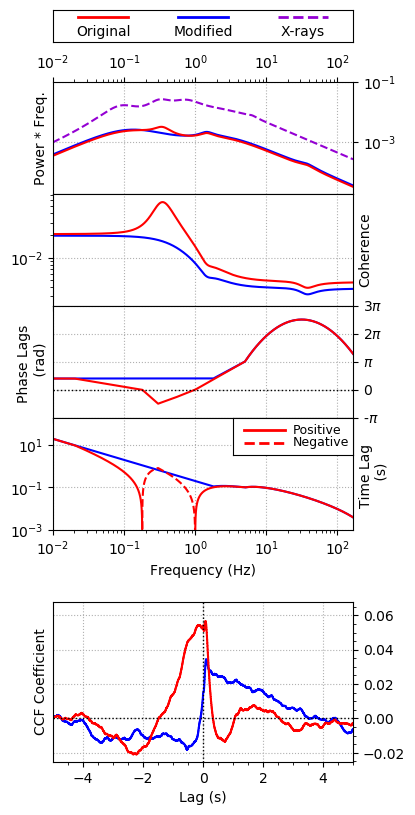}
		\caption{Two simulations of the $i_{s}$ band with X-rays from epoch 6. \textbf{Top:} Input Fourier components. The red lines are a representation of the data as it was seen in Figs. \ref{fig:powerspectra}--\ref{fig:time_lags}, and the blue lines are a modification that removes the QPO and the negative lags from the $i_{s}$ band's Fourier components between 0.02--2\,Hz. \textbf{Bottom:} CCFs made by converting the Fourier components into lightcurves and then cross-correlating the results. CCFs were averaged over multiple 10s segments. Note how the behaviour completely changes between -2 and +3\,s, showing the significance of epoch 6's negative lags over this range.}
		\label{fig:Epoch6_CorrSim}
	\end{figure}

	
	The QPO, along with this different behaviour component, are both strong features in epoch 6. To what magnitude, and in what ways, do they affect epoch 6's cross-correlation (Figs. \ref{fig:10s_CCFs} \& \ref{fig:2s_CCFs})? To find out, we simulated an approximation of the Fourier components of the X-rays and $i_s$ band of epoch 6, creating a lightcurve for each from these components, and then cross-correlated them. We then modified the Fourier components to remove both the Lorentzian responsible for the QPO and the negative lags; for the latter, we instead assumed an interpolated flat distribution of 2$\pi$/5 in the phase lags below 2\,Hz. A CCF was made from these lightcurves as well, and the two results (as well as the inputs) are shown in Figure \ref{fig:Epoch6_CorrSim}. 
	
	Significantly, it can be seen how the cross-correlation is entirely different between -2\,s and 3\,s lags, no longer showing the negative correlations unique to epoch 6, nor the positive anti-correlation that is present in epochs 4 and 6. From this, we conclude that the QPO and the negative lags are the primary cause for the oddities we see in the epoch 6 CCF. For more information, including how each component affects the CCF individually and further evidence of the QPO influencing positive as well as negative lags, see Appendix \ref{sec:simulated_fourier_components_extra}.
	
	\section{Discussion} \label{discuss}
	
	Analysis of our results has shown both features that are constant, and ones that are varying in specific ways over the course of the outburst. To summarise our main findings:
	
	\begin{enumerate}
	
	    \item In all epochs, J1820 shows rapid, sub-second red flares, and longer-scale variations that are stronger at longer wavelengths (Fig. \ref{fig:Lightcurves}).
	    
	    \item Over the epochs, an anti-correlation component (stronger at longer wavelengths) around zero lag becomes increasingly significant -- until late into the hard state decline, when it is superseded by a \textit{positive} correlation at negative lags, again stronger at longer wavelengths (Fig. \ref{fig:10s_CCFs}).
	    
	    \item In all epochs, the CCF reveals a sub-second peak in the optical/X-ray correlation function at roughly 0.2\,s. The peak maintains a roughly similar shape over the epochs, but appears to shrink in comparison to other features
	    (Fig. \ref{fig:2s_CCFs}).
	    
	    \item Over time, optical bands become less variable (i.e. decrease in power) at higher frequencies, but the variability/power remains roughly constant at lower frequencies. For X-rays, this relation is inverted, showing an overall decrease of rms$^2$ power at low frequencies. The optical power spectra also have consistently higher rms$^2$ power at longer wavelengths  (Fig. \ref{fig:powerspectra}).
	    
	    
	    \item Coherence at lower frequencies drops as the outburst continues (Fig. \ref{fig:coherence}).
	    
	    \item The phase/time lags are mostly consistent between 1-10\,Hz across epochs. At lower frequencies, they change from being near +$\pi$/4 to being near $\pm\pi$ as the outburst progresses. Epoch 6, however, fits neither of these trends (Fig. \ref{fig:phase_lags}).
	    
	    \item All epochs have an interval between Fourier frequencies 1--5\,Hz where shorter wavelengths have shorter time lags. This behaviour is roughly consistent (aside from in epoch 6), and neither the lag nor separation by wavelength appear to change (Fig. \ref{fig:time_lags}).
	    
	\end{enumerate}
	
	In these observations lies evidence for evolving processes within the system. We will now address several key points and theories based on these observations.

	\subsection{Compact Jet} \label{sec:compact_jet}
	
	Jet activity has already been found in this source (e.g. \citealt{Homan_J1820_2018, Bright_J1820_2020}), and the presence of rapid red variations and a sub-second optical lag that we show in this paper can both result from jet activity \citep{gandhi_elevation_2017}. Radio data \citep{Bright_J1820_2020} show the source to be relatively bright in the radio, and the long-term lightcurve approximates that in hard X-rays (Fig. \ref{fig:timeline}). Meanwhile, \citet{Russell_1820_ATel_2018} presented evidence that the optical emission was likely on the optically-thin tail of synchrotron power-law emission from a jet during April 2018.
	
	However, another interesting phenomenon ties in with this: the wavelength dependence of the sub-second optical lag. In \citet{Paice_1820Letter_2019} we investigated the data shown in epoch 4, and it was first found that a component of the optical emission lagged the X-rays by roughly 170ms. It was also found that this lag was dependent on wavelength; shorter wavelengths lagged less, and longer wavelengths lagged more.
	
	In the previous paper, it was suggested that this feature is emission from a compact jet. In this interpretation, we consider material emitting in hard X-rays close to the compact object from a jet-emitting disk (\citealt{Marcel_AccEject_2019}; though it is also theorised that X-rays may come from the jet itself: \citealt{Markoff_Coronae_2005, Kylafis_JetModel_2008}). A portion of this material is then ejected as a jet; with a fluctuating ejection rate, this does not necessarily lead to a uniform stream along the jet, but instead an outflow that varies in density and/or Lorentz factor over time. We can interpret this as a series of discrete shells of matter; since these shells vary in speed, faster shells can thus collide with earlier, slower shells. When they do, they emit through synchrotron radiation. This is the internal shock model \citep{Jamil_iShocks_2010, Malzac_Shocks_2013, Malzac_InternalShocks_2014}, the development of which has been motivated by research into Gamma-Ray Bursts (GRBs) and Active Galactic Nuclei (AGN) jets \citep{Rees_Meszaros_GammaRay_1994, Beloborodov_GRBInternalShocks_2000, Tchekhovskoy_LaunchingAGNJets_2015}. 
	\citet{Tetarenko_J1820Jet_2021} found that the jet in this source is highly relativistic ($\Gamma$ = 6.81); this would mean that a time delay of 170\,ms corresponds to roughly 5$\times10^{4}$\,km between the X-ray and this synchrotron emitting region.
	
	The energy of this synchrotron emission is dependent upon the variation in the Lorentz factor of the colliding material; a larger gradient produces higher energy dissipation. Collisions between larger gradients also occur closer to the compact object, and thus at shorter time lags. Since the regions close to the compact object are more compact, synchrotron emission from these regions is more self-absorbed  and peaks at shorter wavelength. Thus we see shorter time lags for shorter wavelength. A difference of 20\,ms between $z_s$ and $u_s$ peak lags would, for a highly relativistic jet, correspond to a spatial extent of 6$\times10^{3}$\,km.
	
    With the new observations presented in this paper, we have found that this behaviour is also not only present across \textit{all} our epochs between 1--5\,Hz in Fourier frequency, but it also appears to be fairly \textit{consistent} in that range (with the exception of epoch 6, where a different component has the opposite effect on wavelengths up to 2\,Hz) and is independent of the shape of the X-ray power spectrum.
	
	However, while the behaviour stays more or less consistent, the relative contribution of this process to the overall variability appears to decrease over time; note the decreasing significance of the sub-second peak in Figures \ref{fig:10s_CCFs} \& \ref{fig:2s_CCFs}. We also note the significantly changing phase lags;  \citet{Paice_1820Letter_2019} suggested that $\pm\pi$ phase lags at low (<$\sim$1\,Hz) frequencies could be a sign of Doppler-boosting of a jet in high-inclination systems, which was put forward by \citet{Malzac_GX339_2018}. However, our analysis (see Figure \ref{fig:phase_lags}) now shows that $\pm\pi$ phase lags are not a constant feature of this source, and only appear in the short timescales covered by epochs 3-5. Additionally, the Coherence also decreases over time as the lags change, similar to what has been seen in GX 339-4 \citep[][though note that this compares low hard-state and hard-intermediate-state observations]{Vincentelli_SubSecondVariability_2019}.
	
	Over this same range, the X-ray power spectra at these frequencies also decrease in strength over time, with a sharp decrease between epochs 1 and 2, where there is also a sharp decrease seen in the CCFs. \citet{WijnandsvanderKlis_PowerSpectra_1999} notes that the Lorentzians that can describe the X-ray power spectra move to higher frequency over an outburst, which leads to such decreases in power at low frequencies. This is interpreted as resulting from changes in the source geometry.
	
	
    What do we know of the evolution of the geometry of the source? \citet{Kara_MAXIJ1820Corona_2019} found that the corona appears to shrink over the course of the hard state, based on a model that assumed a disc that extends to the innermost stable circular orbit. In our data, we see a broad anti-correlation, which is more often attributed to a hot flow inside a truncated disc (See Section \ref{sec:hot_flow}). \citet{Zdziarski_GeometryJ1820_2021}, meanwhile, describes a \textit{radially} decreasing corona and also features a truncated disc, inside which is a hot flow. In either scenario, an increasingly compact corona could mean that the X-ray emission from it would contribute less to variability at lower frequencies, and would also correspond with a decrease in the significance of the jet component (because both the corona and the jet are linked through fluctuations in accretion power, which heat the X-ray emitting corona and power the jet; thus, changes in one indicate changes in the other; \citealt{Markoff_Coronae_2005}). Overall, the corona becoming more compact would, by itself and its effect on the jet, explain several of features that we see.

	
	The optical QPO could also be explained by a precessing jet. This geometrically-based interpretation has the corona, which is connected to the jet, precessing in such a way that it creates variability in the lightcurves. This has been demonstrated in, e.g., \citet{Liska_PrecessingMHD_2018}, though is still a matter of debate (see, e.g., \citealt{IngramMotta_QPOReview_2020, MarcelNeilsen_QPOSupersonic_2021}). The QPO may also contribute to the anti-correlation around zero lag in epochs 3--5; the high coherence at the QPO frequency would mean that smooth oscillations would be seen in the CCF, and the anti-correlation occurs between the QPO correlation peaks of -3\,s and 5\,s (also worth noting is that the QPO is stronger at longer wavelengths, a fact which is also true of the anti-correlation). It is thus feasible that the QPO contributes to the strength of the anti-correlation, though it need not necessarily be the sole cause of it (for instance, an anti-correlation at negative lags is present in epochs 1 and 2, when no QPO could be seen in the optical power spectra).

	\subsection{Truncated Disc and Inner Accretion Flow} \label{sec:hot_flow}
	
	Is the disc truncated, and if so, does its inner radius evolve? \citet{Buisson_NuSTARJ1820_1_2019} noted that, using relativistic reflection models, the inner edge of the accretion disc appears to remain steady and close to ISCO during most of the hard state; however, \citet{Zdziarski_GeometryJ1820_2021} noted the inner radius of the disc being much more truncated, and evolving over time, moving inwards overall (though perhaps in a stochastic fashion). Likewise, \citet{DeMarco_J1820InnerFlow_2021} reported a truncated disc moving closer to the black hole as the hard state evolved.
	
	If a disc's innermost radius is recessed from the black hole, then there is potential for a hot accretion flow to form. Both the observed fast UV/X-ray timing \citep{Kajava_Dips_2019} and low optical polarization \citep{Veledina_Polarisation_2019} properties can be explained in terms of this (optically thin, geometrically thick) hot inner flow \citep{EardleyLightman_TwoTempAccDisc_1975, Ichimaru_BimodalAccDiscs_FirstADAF_1977, Narayan_Yi_ADAF_1994}.
	
	Our observations show several features that could indicate this as a significant process. For instance, the CCFs in Figure \ref{fig:10s_CCFs} show the presence of an anti-correlation in several of the epochs.

	The anti-correlation can be expected if the hot flow broadband spectrum has a pivoting point, e.g. if an increase of mass accretion rate leads to an increase in X-ray luminosity, at the same time causing higher synchrotron self-absorption within the flow (as a result of higher electron number density), thus leading to a drop in optical emission \citep{veledina_synchrotron_2011}. In this scenario, the variability amplitude is higher at energies further away from the pivoting point, hence we expect to have stronger variability at longer wavelengths, as observed (Fig.~\ref{fig:Lightcurves}).
	
	
	In order to explain the complex anti-correlations at both positive and negative lags in epochs 4--6 in terms of the hot flow scenario, one needs to have two sources of both X-ray and optical emission \citep{veledina_swiftj1753ccfs_2017}; X-rays would be produced by disc and synchrotron Comptonization, and optical by synchrotron emission in the hot flow and irradiated disc emission. These features may appear in the spectrum close to the state transition.
	The natural expectation of such scenario is the different shape of the correlation with soft and hard X-rays, which we indeed see (more details in Section~\ref{sec:epoch6ccfs}).

	The presence of a simultaneous QPO at X-ray and optical wavelengths is another expectation of the hot flow scenario \citep{Veledina_Precession_2013}, which seems to be confirmed by our data from epochs 4 and 6. A correlated QPO can significantly alter the shape of the CCF \citep[see, e.g.][]{veledina_swiftj1753ccfs_2017} and can potentially explain some features of the epoch 6 CCF (see Section~\ref{sec:epoch6} and Fig.~\ref{fig:Epoch6_CorrSim} for more discussion). On the other hand, the amplitude of phase lag at the QPO frequency, $\sim-\pi/2$, is not consistent with the expectation of the linear theory, which suggests either 0 or $\pi$ depending on the system orientation (\citealt{Veledina_Precession_2013}; though it is worth noting that the QPO phase lag was closer to $\pm\pi$ at earlier epochs). Furthermore, the lag at the QPO frequency can be altered by the aperiodic component -- however, quantitative conclusions on this possibility can only be drawn from dedicated simulations, which are beyond our present scope. Alternatively, if we assume that the true phase lag is positive (i.e., shifted by $2\pi$), the reprocessing signal can contribute to the QPO \citep{veledina_reprocessing_2015}: 0.3~Hz if within the range of frequencies at which the reprocessed QPO is not smeared out by the light travel delays.
	
	The hot accretion flow scenario can explain most of the changing components in the CCF from different epochs, but not the steady narrow peaks at sub-second lags. The fast optical correlation, most probably coming from a separate emission component, has to be added to the hot flow contribution to get the overall CCF shape consistent with the data.

	\subsection{Epoch 6 and the Emergence of Superhumps} \label{sec:Superhump}

	
	Towards the end of the hard state, a superhump modulation at a period of $\sim$0.7\,days was first reported in the optical light-curve of J1820 by \citet{Patterson_SuperhumpATel_2018}, and then later expanded upon in \citet{Patterson_Superhump_2019}. This signal appeared around day 87 (MJD 58275), with post-hoc analysis revealing that it may have appeared as early as MJD 58272.	Epoch 6 took place on MJD 58276.2, very soon after the superhump appeared. Considering the times of maximum light noted in \citet{Patterson_Superhump_2019}, and assuming a period of 0.7\,days, a maximum occurred at MJD 58276.23, essentially concurrent with epoch 6.
	
	To date, there have been very few studies into the effect of superhumps on optical/X-ray correlations. Given that superhumps are considered to be a property of the {\it outer} disc (see \citealt{WhitehurstKing_Superhumps_1991}), the timescales involved will correspond to the light-travel time to the disc's tidal radius, which for J1820 will be $\sim$10s, and hence any correlated variations are likely to be heavily smeared, compared to the timescales being studied here.  Actually, optical/X-ray CCFs were constructed for the black hole LMXB Swift J1753.5--0127, and were found to be independent of the superhump period present in that system (see Section 3 of \citealt{durant_swift_2008}, and note that ``orbital-like modulation'' refers to superhumps). However, there have been no studies that examined phase lags in this scenario. Thus, this avenue of research would be valuable in investigating whether or not they contribute to the features we see in epoch 6, and, by extension, might be affecting the optical/X-ray correlations and variability of LMXB systems as a whole.  Further studies of J1820's superhump properties can be found in Thomas et al. (2021; Subm.).

	\subsection{A Combined Jet and Hot Flow Model} \label{sec:Model}
	
	Let us now link our findings to the various models presented. The source shows repeated rapid red flares, and a sub-second optical/X-ray correlation that has a larger lag at longer wavelengths. The components dominating the correlation at low frequencies change as the hard state evolves; the X-ray power spectra and the optical/X-ray coherence both decrease at these frequencies, and the phase lags move towards $\pm\pi$. The source also becomes softer over time, and the sub-second lag in the cross-correlations becomes less significant. Meanwhile, the X-ray power and the coherence at higher frequencies remains static.
	
	We do not find that the donor star is an explanation for our features; while the star could theoretically produce a correlated component at positive lags in our CCFs through X-ray heating and reprocessing, combining mass and orbital period estimates from \citet{Atri_J1820Parallax_2020} and \citet{Torres_J1820Mass_2019} with Kepler's third law gives the distance between the compact object and donor to be $\sim$16\,light-seconds, and the effect in the lags would likely vary between epochs as we observe different phases, in disagreement with either the smooth evolution or constant nature of the correlated components we see. However, given the high system inclination ($\sim75\degr$, \citealt{Torres_J1820Mass_2019}), the shortest delays between X-ray and (reprocessed) optical photons from the near-side of the disc are expected to be about $\sim0.5$\,s, with some additional smearing to longer lags due to light travel times across the face of the disc. Hence, it is possible that X-ray reprocessing off the accretion disc could be significant to the variability; this can be tested in future by comparing these results to similar soft-state observations, where the illuminating component should be more dominant. 

    In all, we suggest a two-component model; one correlated, and the other anti-correlated. The correlated component we ascribe to a compact jet, which becomes less significant over time. The anti-correlated component, meanwhile, we ascribe to a hot flow, which remains static.

    A jet as the correlated component would explain the red flares, the optical/X-ray sub-second correlation \citep{gandhi_elevation_2017}, and the larger lag at longer wavelengths \citep{Malzac_Shocks_2013, Malzac_InternalShocks_2014}. X-rays coming from the inflow would contribute more to the X-ray variability at the lowest (<0.1\,Hz) and the highest (>1\,Hz) frequencies. If the corona is contracting (Evidenced either by a change in the vertical extent, as in \citealt{Kara_MAXIJ1820Corona_2019}, or a change in the radial extent and a decreasing disc truncation radius, as in \citealt{Zdziarski_GeometryJ1820_2021}), the variability of hard X-rays from that corona would decrease at lower frequencies, as would the optical/X-ray coherence over the same range -- while the jet, closely linked to the corona, would also decrease in significance, leading to the decline of the sub-second correlation.

    The latter, anti-correlated, component we ascribe to the hot flow. This component stays mostly static, and thus, relatively, contributes more to the overall variability as the jet declines in significance. A hot flow scenario could feasibly also explain the QPO that we see in the data. The hot flow does not appear to increase in significance -- note that the coherence does not increase.
    
    
    \subsubsection{Beyond the Jet and Hot Flow}
    
    \citet{MunozDarias_PCygJ1820_2019} and \citet{SanchezSierrasMunozDarias_NIRWindsJ1820_2020} reported the detection of optical and near-infrared winds respectively in J1820. The effect of winds on optical/X-ray timing correlations has not yet been explored in depth, however they would occur on similar timescales to those studied here. V404 Cyg is a similar system to J1820 (albeit with a much longer orbital period of 6.47 days and thus a larger physical scale; \citealt{Casares_V404_1992}); in that source, the wind launching zone was found to be on the order of a few $\times10^5$\,km \citep{MunozDarias_V404Winds_2016}, or about 0.5\,lightseconds. For a source inclination of 75$\degr$ \citep{Torres_J1820Mass_2019}, and using eq.\,(4) in \cite{poutanen_impact_2002}, we get minimum lags on the order of 0.01\,s, so contribution of the wind to the CCF timescales that we probe is feasible from a timing standpoint.
    
    However, the shallowness of the P Cygni absorption feature (1--2 percent below the continuum level, \citealt{MunozDarias_PCygJ1820_2019}) implies that the wind is optically thin, which would mean that there would be minimal reprocessed emission due to the wind. Further investigation into this possibility would require better data on the optical depth and the ionization of the wind, combined with simulations.

	\section{Conclusions}

    We have presented analysis of optical and X-ray lightcurves from the black hole Low-Mass X-ray Binary (LMXB) MAXI J1820+070 over the course of roughly 80 days. In doing so, we show an evolving Cross Correlation Function (CCF) at longer ($\sim$10\,second) timescales, a consistent sub-second correlation, and various changes in the Fourier components, including differences between different optical wavelengths.
	
	This paper thus shows both the dynamic and static nature of LMXBs, even over a single outburst. The shifting of phase lags at lower frequencies, the slowly climbing photon index, and the increasingly significant anti-correlation shows how the coherent components can change on a timescale of weeks. Meanwhile, the constant nature of the correlation at sub-second lags, mid-frequency time lags, and rapid red flares in the lightcurves show that other components are more stable, and can be present with broadly static properties more than two months apart. Additionally, it shows how a Quasi-Periodic Oscillation (QPO), travelling upwards through the Fourier frequencies, can change the resultant lags and correlation features.
	
	We discuss our findings in terms of two synchrotron-emitting components -- a correlated jet and an anti-correlated hot flow -- as major contributors to the overall variability. If we allow for the jet to dominate at the lowest (<0.1\,Hz) and the highest (>1\,Hz) frequencies, and the hot flow to dominate in between, the interaction of these components can create the features we observe in several epochs. 
	
	A correlated component at negative lags can be seen in several epochs. Fourier analysis showed this component to be related to the frequency of a QPO in both the optical and X-ray lightcurves, previously reported in X-rays by \citet{Stiele_Kong_J1820_Evolution_2020}. The lightcurves are consistently coherent at these frequencies, with greater coherence (and thus correlation) at longer wavelengths. As the QPO increases in frequency over the outburst, the lag also evolves, becoming less negative. We note that, due to the periodic nature of the QPO, this negative lag could easily be a Fourier artefact, and the true lag is positive, with X-ray variability leading optical by several seconds.
	
	Epoch 6 shows us features that are more difficult to understand. Between 0.08 to 2.5\,Hz, there is some component that causes a drop in optical/X-ray phase lags. This component is more significant at longer wavelengths, and the lags become negative in most bands. The QPO mentioned earlier is in the middle of these frequencies, but there is no indication as to whether it is related or not. 
	Further observations of LMXBs close to the intermediate state would be highly desirable to investigate this.
	
	The evolution of the optical/X-ray correlations over the course of an LMXB's outburst remains an area rich with possibility for new discoveries. This paper highlights the fact that further, more frequent investigations of an LMXB over its hard state (and, ideally, over the transition to the soft state) would be invaluable in further decoding the shifting phenomena inside these sources.
	
	
	

	\section*{Data Availability}

    The NICER data underlying this article are available in the HEASARC Data Archive (\href{https://heasarc.gsfc.nasa.gov/docs/archive.html}{https://heasarc.gsfc.nasa.gov/docs/archive.html}). The ULTRACAM and HiPERCAM data will be shared on reasonable request to the corresponding author. The Swift data from Fig. \ref{fig:timeline} are available from the Swift Archive (\href{https://www.swift.ac.uk/archive/}{https://www.swift.ac.uk/archive/}). The AMI-LA data from Fig. \ref{fig:timeline} are available from \citet{Bright_J1820_2020} (\href{https://www.nature.com/articles/s41550-020-1023-5}{https://www.nature.com/articles/s41550-020-1023-5}).

	\section*{Acknowledgements}

	We acknowledge support from STFC and a UGC-UKIERI Thematic Partnership.
	We would like to thank the anonymous referee for their helpful comments. We would also like to thank Joe Bright, Piergiorgio Casella, Rob Fender, Adam Ingram, Sera Markoff, Sara Motta, Tom Russell, Gregory Sivakoff, and Alex Tetarenko for their helpful conversations. We thank our ULTRACAM observers Paul Chote, Martin Dyer, and Anna Pala. We also thank Keith Gendreau, Zaven Arzoumanian, and the rest of the NICER team for their assistance in coordinating observations.
	
	JAP is part supported by a University of Southampton Central VC Scholarship, and thanks D Ashton for spectral timing help, as well as A Stevens and D Huppenkothen for help with the Stingray software. 
	TS thanks the Spanish Ministry of Economy and Competitiveness (MINECO; grant AYA2017-83216).
	KR acknowledges funding from the European Research Council (ERC) under the European Union’s Horizon 2020 research and innovation programme (grant agreement No. 694745).
    AV acknowledges the Academy of Finland grant 309308 and the International Space Science Institute (ISSI) in Bern, Switzerland for support.
	This work was supported by the Programme National des Hautes Energies of CNRS/INSU with INP and IN2P3, co-funded by CEA and CNES. 
    HiPERCAM and VSD were funded by the European Research Council (FP/2007–2013) under ERC-2013-ADG grant agreement no. 340040. ULTRACAM and VSD are funded by the STFC.
    FMV acknowledges support from STFC under grant ST/R000638/1.
    
	HiPERCAM observations were made with the GTC telescope (Spanish Observatorio del Roque de los Muchachos, Instituto de Astrof\'\i{}sica de Canarias), under Director's Discretionary Time. SMARTNet helped to coordinate observations. We have made use of software and web tools from the High Energy Astrophysics Science Archive Research Center (HEASARC), and made use of data and the 'Build XRT Products' tool supplied by the UK Swift Science Data Centre at the University of Leicester.

	\bibliographystyle{mnras}
	\bibliography{main}

\begin{thebibliography}{}
\makeatletter
\relax
\def\mn@urlcharsother{\let\do\@makeother \do\$\do\&\do\#\do\^\do\_\do\%\do\~}
\def\mn@doi{\begingroup\mn@urlcharsother \@ifnextchar [ {\mn@doi@}
  {\mn@doi@[]}}
\def\mn@doi@[#1]#2{\def\@tempa{#1}\ifx\@tempa\@empty \href
  {http://dx.doi.org/#2} {doi:#2}\else \href {http://dx.doi.org/#2} {#1}\fi
  \endgroup}
\def\mn@eprint#1#2{\mn@eprint@#1:#2::\@nil}
\def\mn@eprint@arXiv#1{\href {http://arxiv.org/abs/#1} {{\tt arXiv:#1}}}
\def\mn@eprint@dblp#1{\href {http://dblp.uni-trier.de/rec/bibtex/#1.xml}
  {dblp:#1}}
\def\mn@eprint@#1:#2:#3:#4\@nil{\def\@tempa {#1}\def\@tempb {#2}\def\@tempc
  {#3}\ifx \@tempc \@empty \let \@tempc \@tempb \let \@tempb \@tempa \fi \ifx
  \@tempb \@empty \def\@tempb {arXiv}\fi \@ifundefined
  {mn@eprint@\@tempb}{\@tempb:\@tempc}{\expandafter \expandafter \csname
  mn@eprint@\@tempb\endcsname \expandafter{\@tempc}}}

\bibitem[\protect\citeauthoryear{{Adachi} et~al.,}{{Adachi}
  et~al.}{2020}]{ATel13502}
{Adachi} R.,  et~al., 2020, The Astronomer's Telegram, \href
  {https://ui.adsabs.harvard.edu/abs/2020ATel13502....1A} {13502, 1}

\bibitem[\protect\citeauthoryear{{Atri} et~al.,}{{Atri}
  et~al.}{2020}]{Atri_J1820Parallax_2020}
{Atri} P.,  et~al., 2020, \mn@doi [\mnras] {10.1093/mnrasl/slaa010}, \href
  {https://ui.adsabs.harvard.edu/abs/2020MNRAS.493L..81A} {493, L81}

\bibitem[\protect\citeauthoryear{{Baglio}, {Russell}  \& {Lewis}}{{Baglio}
  et~al.}{2018}]{ATel11418}
{Baglio} M.~C.,  {Russell} D.~M.,   {Lewis} F.,  2018, The Astronomer's
  Telegram, \href {http://adsabs.harvard.edu/abs/2018ATel11418....1B} {11418}

\bibitem[\protect\citeauthoryear{{Bahramian}, {Strader}  \& {Dage}}{{Bahramian}
  et~al.}{2018}]{ATel11424}
{Bahramian} A.,  {Strader} J.,   {Dage} K.,  2018, The Astronomer's Telegram,
  \href {https://ui.adsabs.harvard.edu/abs/2018ATel11424....1B} {11424, 1}

\bibitem[\protect\citeauthoryear{{Belloni} \& {Hasinger}}{{Belloni} \&
  {Hasinger}}{1990a}]{BelloniVariability1990}
{Belloni} T.,  {Hasinger} G.,  1990a, A\&A, \href
  {http://adsabs.harvard.edu/abs/1990A%26A...227L..33B} {227, L33}

\bibitem[\protect\citeauthoryear{{Belloni} \& {Hasinger}}{{Belloni} \&
  {Hasinger}}{1990b}]{Belloni_Hasinger_Aperiodic_1990}
{Belloni} T.,  {Hasinger} G.,  1990b, \aap, \href
  {https://ui.adsabs.harvard.edu/abs/1990A&A...230..103B} {230, 103}

\bibitem[\protect\citeauthoryear{{Beloborodov}}{{Beloborodov}}{2000}]{Beloborodov_GRBInternalShocks_2000}
{Beloborodov} A.~M.,  2000, \mn@doi [\apjl] {10.1086/312830}, \href
  {https://ui.adsabs.harvard.edu/abs/2000ApJ...539L..25B} {539, L25}

\bibitem[\protect\citeauthoryear{{Bright}, {Fender}  \& {Motta}}{{Bright}
  et~al.}{2018}]{ATel11420}
{Bright} J.,  {Fender} R.,   {Motta} S.,  2018, The Astronomer's Telegram,
  \href {http://adsabs.harvard.edu/abs/2018ATel11420....1B} {11420}

\bibitem[\protect\citeauthoryear{{Bright} et~al.,}{{Bright}
  et~al.}{2020}]{Bright_J1820_2020}
{Bright} J.~S.,  et~al., 2020, \mn@doi [Nature Astronomy]
  {10.1038/s41550-020-1023-5}, \href
  {https://ui.adsabs.harvard.edu/abs/2020NatAs...4..697B} {4, 697}

\bibitem[\protect\citeauthoryear{{Buisson} et~al.,}{{Buisson}
  et~al.}{2019}]{Buisson_NuSTARJ1820_1_2019}
{Buisson} D.~J.~K.,  et~al., 2019, \mn@doi [\mnras] {10.1093/mnras/stz2681},
  \href {https://ui.adsabs.harvard.edu/abs/2019MNRAS.490.1350B} {490, 1350}

\bibitem[\protect\citeauthoryear{{Casares}, {Charles}  \& {Naylor}}{{Casares}
  et~al.}{1992}]{Casares_V404_1992}
{Casares} J.,  {Charles} P.~A.,   {Naylor} T.,  1992, \mn@doi [\nat]
  {10.1038/355614a0}, \href
  {https://ui.adsabs.harvard.edu/abs/1992Natur.355..614C} {355, 614}

\bibitem[\protect\citeauthoryear{Casella et~al.,}{Casella
  et~al.}{2010}]{casella_fast_2010}
Casella P.,  et~al., 2010, \mn@doi [Monthly Notices of the Royal Astronomical
  Society: Letters] {10.1111/j.1745-3933.2010.00826.x}, 404, L21

\bibitem[\protect\citeauthoryear{Corral-Santana et~al.,}{Corral-Santana
  et~al.}{2016}]{corral-santana_blackcat:_2016}
Corral-Santana J.~M.,  et~al., 2016, \mn@doi [\aap]
  {10.1051/0004-6361/201527130}, 587, A61

\bibitem[\protect\citeauthoryear{{De Marco}, {Zdziarski}, {Ponti}, {Migliori},
  {Belloni}, {Segovia Otero}, {Dzie{\l}ak}  \& {Lai}}{{De Marco}
  et~al.}{2021}]{DeMarco_J1820InnerFlow_2021}
{De Marco} B.,  {Zdziarski} A.~A.,  {Ponti} G.,  {Migliori} G.,  {Belloni}
  T.~M.,  {Segovia Otero} A.,  {Dzie{\l}ak} M.,   {Lai} E.~V.,  2021, arXiv
  e-prints, \href {https://ui.adsabs.harvard.edu/abs/2021arXiv210207811D} {p.
  arXiv:2102.07811}

\bibitem[\protect\citeauthoryear{{Del Santo} \& {Segreto}}{{Del Santo} \&
  {Segreto}}{2018}]{ATel11427}
{Del Santo} M.,  {Segreto} A.,  2018, The Astronomer's Telegram, \href
  {https://ui.adsabs.harvard.edu/abs/2018ATel11427....1D} {11427, 1}

\bibitem[\protect\citeauthoryear{{Denisenko}}{{Denisenko}}{2018}]{ATel11400}
{Denisenko} D.,  2018, The Astronomer's Telegram, \href
  {http://adsabs.harvard.edu/abs/2018ATel11400....1D} {11400}

\bibitem[\protect\citeauthoryear{Dhillon et~al.,}{Dhillon
  et~al.}{2007}]{dhillon_ultracam:_2007}
Dhillon V.~S.,  et~al., 2007, \mn@doi [Monthly Notices of the Royal
  Astronomical Society] {10.1111/j.1365-2966.2007.11881.x}, 378, 825

\bibitem[\protect\citeauthoryear{{Dhillon} et~al.,}{{Dhillon}
  et~al.}{2018}]{Dhillon_First_2018}
{Dhillon} V.,  et~al., 2018, in Ground-based and Airborne Instrumentation for
  Astronomy VII. p. 107020L (\mn@eprint {arXiv} {1807.00557}),
  \mn@doi{10.1117/12.2312041}

\bibitem[\protect\citeauthoryear{{Done}, {Gierli{\'n}ski}  \& {Kubota}}{{Done}
  et~al.}{2007}]{Done_EverythingAccretion_2007}
{Done} C.,  {Gierli{\'n}ski} M.,   {Kubota} A.,  2007, \mn@doi [\aapr]
  {10.1007/s00159-007-0006-1}, \href
  {https://ui.adsabs.harvard.edu/abs/2007A&ARv..15....1D} {15, 1}

\bibitem[\protect\citeauthoryear{Durant et~al.,}{Durant
  et~al.}{2008}]{durant_swift_2008}
Durant M.,  et~al., 2008, \mn@doi [The Astrophysical Journal Letters]
  {10.1086/590906}, 682, L45

\bibitem[\protect\citeauthoryear{{Durant} et~al.,}{{Durant}
  et~al.}{2011}]{Durant_OptXCCFs_2011}
{Durant} M.,  et~al., 2011, \mn@doi [\mnras]
  {10.1111/j.1365-2966.2010.17604.x}, \href
  {https://ui.adsabs.harvard.edu/abs/2011MNRAS.410.2329D} {410, 2329}

\bibitem[\protect\citeauthoryear{{Eardley}, {Lightman}  \& {Shapiro}}{{Eardley}
  et~al.}{1975}]{EardleyLightman_TwoTempAccDisc_1975}
{Eardley} D.~M.,  {Lightman} A.~P.,   {Shapiro} S.~L.,  1975, \mn@doi [\apjl]
  {10.1086/181871}, \href
  {https://ui.adsabs.harvard.edu/abs/1975ApJ...199L.153E} {199, L153}

\bibitem[\protect\citeauthoryear{Eastman, Siverd  \& Gaudi}{Eastman
  et~al.}{2010}]{eastman_achieving_2010}
Eastman J.,  Siverd R.,   Gaudi B.~S.,  2010, \mn@doi [Publications of the
  Astronomical Society of the Pacific] {10.1086/655938}, 122, 935

\bibitem[\protect\citeauthoryear{{Evans} et~al.,}{{Evans}
  et~al.}{2009}]{Evans_SwiftXRTSpectra_2009}
{Evans} P.~A.,  et~al., 2009, \mn@doi [\mnras]
  {10.1111/j.1365-2966.2009.14913.x}, \href
  {https://ui.adsabs.harvard.edu/abs/2009MNRAS.397.1177E} {397, 1177}

\bibitem[\protect\citeauthoryear{{Fabian} et~al.,}{{Fabian}
  et~al.}{2020}]{Fabian_PlungeEmission_2020}
{Fabian} A.~C.,  et~al., 2020, \mn@doi [\mnras] {10.1093/mnras/staa564}, \href
  {https://ui.adsabs.harvard.edu/abs/2020MNRAS.493.5389F} {493, 5389}

\bibitem[\protect\citeauthoryear{{Gaia Collaboration}, {Brown}, {Vallenari},
  {Prusti}, {de Bruijne}, {Babusiaux}  \& {Biermann}}{{Gaia Collaboration}
  et~al.}{2020}]{gaia_EDR3_2020}
{Gaia Collaboration} {Brown} A.~G.~A.,  {Vallenari} A.,  {Prusti} T.,  {de
  Bruijne} J.~H.~J.,  {Babusiaux} C.,   {Biermann} M.,  2020, arXiv e-prints,
  \href {https://ui.adsabs.harvard.edu/abs/2020arXiv201201533G} {p.
  arXiv:2012.01533}

\bibitem[\protect\citeauthoryear{{Gandhi} et~al.,}{{Gandhi}
  et~al.}{2008}]{Gandhi_Correlations_2008}
{Gandhi} P.,  et~al., 2008, \mn@doi [\mnras]
  {10.1111/j.1745-3933.2008.00529.x}, \href
  {https://ui.adsabs.harvard.edu/abs/2008MNRAS.390L..29G} {390, L29}

\bibitem[\protect\citeauthoryear{Gandhi et~al.,}{Gandhi
  et~al.}{2010}]{gandhi_rapid_2010}
Gandhi P.,  et~al., 2010, \mn@doi [Monthly Notices of the Royal Astronomical
  Society] {10.1111/j.1365-2966.2010.17083.x}, 407, 2166

\bibitem[\protect\citeauthoryear{Gandhi et~al.,}{Gandhi
  et~al.}{2016}]{gandhi_furiously_2016}
Gandhi P.,  et~al., 2016, \mn@doi [Monthly Notices of the Royal Astronomical
  Society] {10.1093/mnras/stw571}, 459, 554

\bibitem[\protect\citeauthoryear{Gandhi et~al.,}{Gandhi
  et~al.}{2017}]{gandhi_elevation_2017}
Gandhi P.,  et~al., 2017, \mn@doi [Nature Astronomy]
  {10.1038/s41550-017-0273-3}, 1, 859

\bibitem[\protect\citeauthoryear{{Gandhi}, {Rao}, {Johnson}, {Paice}  \&
  {Maccarone}}{{Gandhi} et~al.}{2019}]{Gandhi_GaiaDR2_2019}
{Gandhi} P.,  {Rao} A.,  {Johnson} M. A.~C.,  {Paice} J.~A.,   {Maccarone}
  T.~J.,  2019, \mn@doi [\mnras] {10.1093/mnras/stz438}, \href
  {https://ui.adsabs.harvard.edu/abs/2019MNRAS.485.2642G} {485, 2642}

\bibitem[\protect\citeauthoryear{{Gendreau} et~al.,}{{Gendreau}
  et~al.}{2016}]{GendreauArzoumanian_NICER_2016}
{Gendreau} K.~C.,  et~al., 2016, in {den Herder} J.-W.~A.,  {Takahashi} T.,
  {Bautz} M.,  eds,  Society of Photo-Optical Instrumentation Engineers (SPIE)
  Conference Series Vol. 9905, Space Telescopes and Instrumentation 2016:
  Ultraviolet to Gamma Ray. p. 99051H, \mn@doi{10.1117/12.2231304}

\bibitem[\protect\citeauthoryear{{Hambsch}, {Ulowetz}, {Vanmunster}, {Cejudo}
  \& {Patterson}}{{Hambsch} et~al.}{2019}]{ATel13014}
{Hambsch} J.,  {Ulowetz} J.,  {Vanmunster} T.,  {Cejudo} D.,   {Patterson} J.,
  2019, The Astronomer's Telegram, \href
  {https://ui.adsabs.harvard.edu/abs/2019ATel13014....1H} {13014, 1}

\bibitem[\protect\citeauthoryear{{Henden}, {Levine}, {Terrell}  \&
  {Welch}}{{Henden} et~al.}{2015}]{Henden_APASS_2015}
{Henden} A.~A.,  {Levine} S.,  {Terrell} D.,   {Welch} D.~L.,  2015, in
  American Astronomical Society Meeting Abstracts \#225. p. 336.16

\bibitem[\protect\citeauthoryear{{Homan} et~al.,}{{Homan}
  et~al.}{2018}]{Homan_J1820_2018}
{Homan} J.,  et~al., 2018, The Astronomer's Telegram, \href
  {https://ui.adsabs.harvard.edu/abs/2018ATel11576....1H} {11576}

\bibitem[\protect\citeauthoryear{{Huppenkothen} et~al.,}{{Huppenkothen}
  et~al.}{2019}]{Huppenkothen_Stingray_2019}
{Huppenkothen} D.,  et~al., 2019, \mn@doi [\apj] {10.3847/1538-4357/ab258d},
  \href {https://ui.adsabs.harvard.edu/abs/2019ApJ...881...39H} {881, 39}

\bibitem[\protect\citeauthoryear{{Ichimaru}}{{Ichimaru}}{1977}]{Ichimaru_BimodalAccDiscs_FirstADAF_1977}
{Ichimaru} S.,  1977, \mn@doi [\apj] {10.1086/155314}, \href
  {https://ui.adsabs.harvard.edu/abs/1977ApJ...214..840I} {214, 840}

\bibitem[\protect\citeauthoryear{{Imamura}, {Steiman-Cameron}  \&
  {Middleditch}}{{Imamura} et~al.}{1987}]{Imamura_1.13msGX339_1987}
{Imamura} J.~N.,  {Steiman-Cameron} T.~Y.,   {Middleditch} J.,  1987, \mn@doi
  [\apjl] {10.1086/184841}, \href
  {https://ui.adsabs.harvard.edu/abs/1987ApJ...314L..11I} {314, L11}

\bibitem[\protect\citeauthoryear{{Ingram} \& {Motta}}{{Ingram} \&
  {Motta}}{2020}]{IngramMotta_QPOReview_2020}
{Ingram} A.,  {Motta} S.,  2020, arXiv e-prints, \href
  {https://ui.adsabs.harvard.edu/abs/2020arXiv200108758I} {p. arXiv:2001.08758}

\bibitem[\protect\citeauthoryear{{Jamil}, {Fender}  \& {Kaiser}}{{Jamil}
  et~al.}{2010}]{Jamil_iShocks_2010}
{Jamil} O.,  {Fender} R.~P.,   {Kaiser} C.~R.,  2010, \mn@doi [\mnras]
  {10.1111/j.1365-2966.2009.15652.x}, \href
  {https://ui.adsabs.harvard.edu/abs/2010MNRAS.401..394J} {401, 394}

\bibitem[\protect\citeauthoryear{{Kajava} et~al.,}{{Kajava}
  et~al.}{2019}]{Kajava_Dips_2019}
{Kajava} J.~J.~E.,  et~al., 2019, arXiv e-prints, \href
  {https://ui.adsabs.harvard.edu/abs/2019arXiv190606519K} {p. arXiv:1906.06519}

\bibitem[\protect\citeauthoryear{Kanbach et~al.,}{Kanbach
  et~al.}{2001}]{kanbach_correlated_2001}
Kanbach G.,  et~al., 2001, \mn@doi [Nature] {10.1038/35102515}, 414, 180

\bibitem[\protect\citeauthoryear{{Kara} et~al.,}{{Kara}
  et~al.}{2019}]{Kara_MAXIJ1820Corona_2019}
{Kara} E.,  et~al., 2019, \mn@doi [\nat] {10.1038/s41586-018-0803-x}, \href
  {https://ui.adsabs.harvard.edu/abs/2019Natur.565..198K} {565, 198}

\bibitem[\protect\citeauthoryear{{Kawamuro} et~al.,}{{Kawamuro}
  et~al.}{2018}]{ATel11399}
{Kawamuro} T.,  et~al., 2018, The Astronomer's Telegram, \href
  {http://adsabs.harvard.edu/abs/2018ATel11399....1K} {11399}

\bibitem[\protect\citeauthoryear{{Kylafis}, {Papadakis}, {Reig}, {Giannios}  \&
  {Pooley}}{{Kylafis} et~al.}{2008}]{Kylafis_JetModel_2008}
{Kylafis} N.~D.,  {Papadakis} I.~E.,  {Reig} P.,  {Giannios} D.,   {Pooley}
  G.~G.,  2008, \mn@doi [\aap] {10.1051/0004-6361:20079159}, \href
  {https://ui.adsabs.harvard.edu/abs/2008A&A...489..481K} {489, 481}

\bibitem[\protect\citeauthoryear{{Lindegren}}{{Lindegren}}{2020}]{Lindegren_ZeroPoints_2020}
{Lindegren} L.,  2020, \mn@doi [\aap] {10.1051/0004-6361/201936161}, \href
  {https://ui.adsabs.harvard.edu/abs/2020A&A...633A...1L} {633, A1}

\bibitem[\protect\citeauthoryear{{Liska}, {Hesp}, {Tchekhovskoy}, {Ingram},
  {van der Klis}  \& {Markoff}}{{Liska}
  et~al.}{2018}]{Liska_PrecessingMHD_2018}
{Liska} M.,  {Hesp} C.,  {Tchekhovskoy} A.,  {Ingram} A.,  {van der Klis} M.,
  {Markoff} S.,  2018, \mn@doi [\mnras] {10.1093/mnrasl/slx174}, \href
  {https://ui.adsabs.harvard.edu/abs/2018MNRAS.474L..81L} {474, L81}

\bibitem[\protect\citeauthoryear{{Littlefield}}{{Littlefield}}{2018}]{ATel11421}
{Littlefield} C.,  2018, The Astronomer's Telegram, \href
  {https://ui.adsabs.harvard.edu/abs/2018ATel11421....1L} {11421, 1}

\bibitem[\protect\citeauthoryear{{Magnier} et~al.,}{{Magnier}
  et~al.}{2020}]{Magnier_PanSTARRS_2020}
{Magnier} E.~A.,  et~al., 2020, \mn@doi [\apjs] {10.3847/1538-4365/abb82a},
  \href {https://ui.adsabs.harvard.edu/abs/2020ApJS..251....6M} {251, 6}

\bibitem[\protect\citeauthoryear{{Malzac}}{{Malzac}}{2013}]{Malzac_Shocks_2013}
{Malzac} J.,  2013, \mn@doi [\mnras] {10.1093/mnrasl/sls017}, \href
  {https://ui.adsabs.harvard.edu/abs/2013MNRAS.429L..20M} {429, L20}

\bibitem[\protect\citeauthoryear{{Malzac}}{{Malzac}}{2014}]{Malzac_InternalShocks_2014}
{Malzac} J.,  2014, \mn@doi [\mnras] {10.1093/mnras/stu1144}, \href
  {https://ui.adsabs.harvard.edu/abs/2014MNRAS.443..299M} {443, 299}

\bibitem[\protect\citeauthoryear{{Malzac} et~al.,}{{Malzac}
  et~al.}{2018}]{Malzac_GX339_2018}
{Malzac} J.,  et~al., 2018, \mn@doi [\mnras] {10.1093/mnras/sty2006}, \href
  {http://adsabs.harvard.edu/abs/2018MNRAS.480.2054M} {480, 2054}

\bibitem[\protect\citeauthoryear{{Marcel} \& {Neilsen}}{{Marcel} \&
  {Neilsen}}{2021}]{MarcelNeilsen_QPOSupersonic_2021}
{Marcel} G.,  {Neilsen} J.,  2021, \mn@doi [\apj] {10.3847/1538-4357/abcbf9},
  \href {https://ui.adsabs.harvard.edu/abs/2021ApJ...906..106M} {906, 106}

\bibitem[\protect\citeauthoryear{{Marcel} et~al.,}{{Marcel}
  et~al.}{2019}]{Marcel_AccEject_2019}
{Marcel} G.,  et~al., 2019, \mn@doi [\aap] {10.1051/0004-6361/201935060}, \href
  {https://ui.adsabs.harvard.edu/abs/2019A&A...626A.115M} {626, A115}

\bibitem[\protect\citeauthoryear{{Markoff}, {Nowak}  \& {Wilms}}{{Markoff}
  et~al.}{2005}]{Markoff_Coronae_2005}
{Markoff} S.,  {Nowak} M.~A.,   {Wilms} J.,  2005, \mn@doi [\apj]
  {10.1086/497628}, \href
  {https://ui.adsabs.harvard.edu/abs/2005ApJ...635.1203M} {635, 1203}

\bibitem[\protect\citeauthoryear{{Mereminskiy}, {Grebenev}, {Molkov},
  {Zaznobin}, {Khorunzhev}, {Burenin}  \& {Eselevich}}{{Mereminskiy}
  et~al.}{2018}]{ATel11488}
{Mereminskiy} I.~A.,  {Grebenev} S.~A.,  {Molkov} S.~V.,  {Zaznobin} I.~A.,
  {Khorunzhev} G.~A.,  {Burenin} R.~A.,   {Eselevich} M.~V.,  2018, The
  Astronomer's Telegram, \href
  {https://ui.adsabs.harvard.edu/abs/2018ATel11488....1M} {11488, 1}

\bibitem[\protect\citeauthoryear{{Motch}, {Ilovaisky}  \& {Chevalier}}{{Motch}
  et~al.}{1982}]{Motch_GX339_1982}
{Motch} C.,  {Ilovaisky} S.~A.,   {Chevalier} C.,  1982, \aap, \href
  {https://ui.adsabs.harvard.edu/abs/1982A&A...109L...1M} {109, L1}

\bibitem[\protect\citeauthoryear{Motch, Ricketts, Page, Ilovaisky  \&
  Chevalier}{Motch et~al.}{1983}]{motch_simultaneous_1983}
Motch C.,  Ricketts M.~J.,  Page C.~G.,  Ilovaisky S.~A.,   Chevalier C.,
  1983, Astronomy and Astrophysics, 119, 171

\bibitem[\protect\citeauthoryear{{Mu{\~n}oz-Darias} et~al.,}{{Mu{\~n}oz-Darias}
  et~al.}{2016}]{MunozDarias_V404Winds_2016}
{Mu{\~n}oz-Darias} T.,  et~al., 2016, \mn@doi [\nat] {10.1038/nature17446},
  \href {https://ui.adsabs.harvard.edu/abs/2016Natur.534...75M} {534, 75}

\bibitem[\protect\citeauthoryear{{Mu{\~n}oz-Darias} et~al.,}{{Mu{\~n}oz-Darias}
  et~al.}{2019}]{MunozDarias_PCygJ1820_2019}
{Mu{\~n}oz-Darias} T.,  et~al., 2019, \mn@doi [\apjl]
  {10.3847/2041-8213/ab2768}, \href
  {https://ui.adsabs.harvard.edu/abs/2019ApJ...879L...4M} {879, L4}

\bibitem[\protect\citeauthoryear{{Mushotzky} et~al.}{{Mushotzky}
  et~al.}{1993}]{Mushotzky_X-ray_1993}
{Mushotzky} R.~F.,  et~al., 1993, \mn@doi [ARA\&A]
  {10.1146/annurev.astro.31.1.717}, \href
  {http://adsabs.harvard.edu/abs/1993ARA%26A..31..717M} {31, 717}

\bibitem[\protect\citeauthoryear{{Narayan} \& {Yi}}{{Narayan} \&
  {Yi}}{1994}]{Narayan_Yi_ADAF_1994}
{Narayan} R.,  {Yi} I.,  1994, \mn@doi [\apjl] {10.1086/187381}, \href
  {https://ui.adsabs.harvard.edu/abs/1994ApJ...428L..13N} {428, L13}

\bibitem[\protect\citeauthoryear{{Negoro} et~al.,}{{Negoro}
  et~al.}{2018}]{ATel12057}
{Negoro} H.,  et~al., 2018, The Astronomer's Telegram, \href
  {https://ui.adsabs.harvard.edu/abs/2018ATel12057....1N} {12057, 1}

\bibitem[\protect\citeauthoryear{{Pahari} et~al.,}{{Pahari}
  et~al.}{2017}]{Pahari_BWCir_2017}
{Pahari} M.,  et~al., 2017, \mn@doi [\mnras] {10.1093/mnras/stx840}, \href
  {https://ui.adsabs.harvard.edu/abs/2017MNRAS.469..193P} {469, 193}

\bibitem[\protect\citeauthoryear{{Paice} et~al.,}{{Paice}
  et~al.}{2018}]{Paice_J1820_2018}
{Paice} J.~A.,  et~al., 2018, The Astronomer's Telegram, \href
  {https://ui.adsabs.harvard.edu/abs/2018ATel11432....1P} {11432, 1}

\bibitem[\protect\citeauthoryear{{Paice} et~al.,}{{Paice}
  et~al.}{2019}]{Paice_1820Letter_2019}
{Paice} J.~A.,  et~al., 2019, \mn@doi [\mnras] {10.1093/mnrasl/slz148}, \href
  {https://ui.adsabs.harvard.edu/abs/2019MNRAS.490L..62P} {490, L62}

\bibitem[\protect\citeauthoryear{{Patterson}}{{Patterson}}{2019}]{Patterson_Superhump_2019}
{Patterson} J.,  2019, in Proc. 38th Annual Conf. Society for Astronomical
  Sciences, ed. R. K. Buchheim et al. (Rancho Cucamonga, CA: Society for
  Astronomical Sciences), 61, http://www.socastrosci.org/Publications.html

\bibitem[\protect\citeauthoryear{{Patterson} et~al.,}{{Patterson}
  et~al.}{2018}]{Patterson_SuperhumpATel_2018}
{Patterson} J.,  et~al., 2018, The Astronomer's Telegram, \href
  {https://ui.adsabs.harvard.edu/abs/2018ATel11756....1P} {11756}

\bibitem[\protect\citeauthoryear{Poutanen}{Poutanen}{2002}]{poutanen_impact_2002}
Poutanen J.,  2002, \mn@doi [Monthly Notices of the Royal Astronomical Society]
  {10.1046/j.1365-8711.2002.05272.x}, 332, 257

\bibitem[\protect\citeauthoryear{{Rees} \& {Meszaros}}{{Rees} \&
  {Meszaros}}{1994}]{Rees_Meszaros_GammaRay_1994}
{Rees} M.~J.,  {Meszaros} P.,  1994, \mn@doi [\apjl] {10.1086/187446}, \href
  {https://ui.adsabs.harvard.edu/abs/1994ApJ...430L..93R} {430, L93}

\bibitem[\protect\citeauthoryear{{Russell} et~al.,}{{Russell}
  et~al.}{2018}]{Russell_1820_ATel_2018}
{Russell} D.~M.,  et~al., 2018, The Astronomer's Telegram, \href
  {https://ui.adsabs.harvard.edu/abs/2018ATel11533....1R} {11533}

\bibitem[\protect\citeauthoryear{{Sako}, {Ohsawa}, {Ichiki}, {Maehara}, {Morii}
   \& {Tanaka}}{{Sako} et~al.}{2018}]{ATel11426}
{Sako} S.,  {Ohsawa} R.,  {Ichiki} M.,  {Maehara} H.,  {Morii} M.,   {Tanaka}
  M.,  2018, The Astronomer's Telegram, \href
  {https://ui.adsabs.harvard.edu/abs/2018ATel11426....1S} {11426, 1}

\bibitem[\protect\citeauthoryear{{S{\'a}nchez-Sierras} \&
  {Mu{\~n}oz-Darias}}{{S{\'a}nchez-Sierras} \&
  {Mu{\~n}oz-Darias}}{2020}]{SanchezSierrasMunozDarias_NIRWindsJ1820_2020}
{S{\'a}nchez-Sierras} J.,  {Mu{\~n}oz-Darias} T.,  2020, \mn@doi [\aap]
  {10.1051/0004-6361/202038406}, \href
  {https://ui.adsabs.harvard.edu/abs/2020A&A...640L...3S} {640, L3}

\bibitem[\protect\citeauthoryear{{Shidatsu} et~al.,}{{Shidatsu}
  et~al.}{2019}]{Shidatsu_Monitoring_2019}
{Shidatsu} M.,  et~al., 2019, \mn@doi [\apj] {10.3847/1538-4357/ab09ff}, \href
  {https://ui.adsabs.harvard.edu/abs/2019ApJ...874..183S} {874, 183}

\bibitem[\protect\citeauthoryear{{Stiele} \& {Kong}}{{Stiele} \&
  {Kong}}{2020}]{Stiele_Kong_J1820_Evolution_2020}
{Stiele} H.,  {Kong} A.~K.~H.,  2020, \mn@doi [\apj]
  {10.3847/1538-4357/ab64ef}, \href
  {https://ui.adsabs.harvard.edu/abs/2020ApJ...889..142S} {889, 142}

\bibitem[\protect\citeauthoryear{{Tchekhovskoy}}{{Tchekhovskoy}}{2015}]{Tchekhovskoy_LaunchingAGNJets_2015}
{Tchekhovskoy} A.,  2015, {Launching of Active Galactic Nuclei Jets}.
Springer, p.~45, \mn@doi{10.1007/978-3-319-10356-3_3}

\bibitem[\protect\citeauthoryear{{Tetarenko} et~al.,}{{Tetarenko}
  et~al.}{2021}]{Tetarenko_J1820Jet_2021}
{Tetarenko} A.~J.,  et~al., 2021, \mn@doi [\mnras] {10.1093/mnras/stab820},
  \href {https://ui.adsabs.harvard.edu/abs/2021MNRAS.tmp..827T} {}

\bibitem[\protect\citeauthoryear{{Timmer} \& {Koenig}}{{Timmer} \&
  {Koenig}}{1995}]{TimmerKoenig_PowerLawNoise_1995}
{Timmer} J.,  {Koenig} M.,  1995, \aap, \href
  {https://ui.adsabs.harvard.edu/abs/1995A&A...300..707T} {300, 707}

\bibitem[\protect\citeauthoryear{{Tomsick} \& {Homan}}{{Tomsick} \&
  {Homan}}{2019}]{ATel12732}
{Tomsick} J.~A.,  {Homan} J.,  2019, The Astronomer's Telegram, \href
  {https://ui.adsabs.harvard.edu/abs/2019ATel12732....1T} {12732, 1}

\bibitem[\protect\citeauthoryear{{Torres}, {Casares}, {Jim{\'e}nez-Ibarra},
  {Mu{\~n}oz-Darias}, {Armas Padilla}, {Jonker}  \& {Heida}}{{Torres}
  et~al.}{2019}]{Torres_J1820Mass_2019}
{Torres} M.~A.~P.,  {Casares} J.,  {Jim{\'e}nez-Ibarra} F.,  {Mu{\~n}oz-Darias}
  T.,  {Armas Padilla} M.,  {Jonker} P.~G.,   {Heida} M.,  2019, \mn@doi
  [\apjl] {10.3847/2041-8213/ab39df}, \href
  {https://ui.adsabs.harvard.edu/abs/2019ApJ...882L..21T} {882, L21}

\bibitem[\protect\citeauthoryear{{Tucker} et~al.,}{{Tucker}
  et~al.}{2018}]{Tucker_ASASSN18ey_2018}
{Tucker} M.~A.,  et~al., 2018, \mn@doi [\apjl] {10.3847/2041-8213/aae88a},
  \href {https://ui.adsabs.harvard.edu/abs/2018ApJ...867L...9T} {867, L9}

\bibitem[\protect\citeauthoryear{{Ulowetz}, {Myers}  \& {Patterson}}{{Ulowetz}
  et~al.}{2019}]{ATel12567}
{Ulowetz} J.,  {Myers} G.,   {Patterson} J.,  2019, The Astronomer's Telegram,
  \href {https://ui.adsabs.harvard.edu/abs/2019ATel12567....1U} {12567, 1}

\bibitem[\protect\citeauthoryear{{Uttley} et~al.,}{{Uttley}
  et~al.}{2018}]{ATel11423}
{Uttley} P.,  et~al., 2018, The Astronomer's Telegram, \href
  {http://adsabs.harvard.edu/abs/2018ATel11423....1U} {11423}

\bibitem[\protect\citeauthoryear{{Vaughan} \& {Nowak}}{{Vaughan} \&
  {Nowak}}{1997}]{Vaughan_Nowak_1997}
{Vaughan} B.~A.,  {Nowak} M.~A.,  1997, \mn@doi [\apjl] {10.1086/310430}, \href
  {http://adsabs.harvard.edu/abs/1997ApJ...474L..43V} {474, L43}

\bibitem[\protect\citeauthoryear{{Veledina}}{{Veledina}}{2016}]{veledina_twocomp_2016}
{Veledina} A.,  2016, \mn@doi [\apj] {10.3847/0004-637X/832/2/181}, \href
  {https://ui.adsabs.harvard.edu/abs/2016ApJ...832..181V} {832, 181}

\bibitem[\protect\citeauthoryear{{Veledina}}{{Veledina}}{2018}]{veledina_twocomp_interplay_2018}
{Veledina} A.,  2018, \mn@doi [\mnras] {10.1093/mnras/sty2556}, \href
  {https://ui.adsabs.harvard.edu/abs/2018MNRAS.481.4236V} {481, 4236}

\bibitem[\protect\citeauthoryear{Veledina \& Poutanen}{Veledina \&
  Poutanen}{2015}]{veledina_reprocessing_2015}
Veledina A.,  Poutanen J.,  2015, \mn@doi [Monthly Notices of the Royal
  Astronomical Society] {10.1093/mnras/stu2737}, 448, 939

\bibitem[\protect\citeauthoryear{Veledina, Poutanen  \& Vurm}{Veledina
  et~al.}{2011}]{veledina_synchrotron_2011}
Veledina A.,  Poutanen J.,   Vurm I.,  2011, \mn@doi [The Astrophysical Journal
  Letters] {10.1088/2041-8205/737/1/L17}, 737, L17

\bibitem[\protect\citeauthoryear{{Veledina}, {Poutanen}  \& {Vurm}}{{Veledina}
  et~al.}{2013a}]{veledina_accretion_2013}
{Veledina} A.,  {Poutanen} J.,   {Vurm} I.,  2013a, \mn@doi [\mnras]
  {10.1093/mnras/stt124}, \href
  {http://adsabs.harvard.edu/abs/2013MNRAS.430.3196V} {430, 3196}

\bibitem[\protect\citeauthoryear{{Veledina}, {Poutanen}  \&
  {Ingram}}{{Veledina} et~al.}{2013b}]{Veledina_Precession_2013}
{Veledina} A.,  {Poutanen} J.,   {Ingram} A.,  2013b, \mn@doi [\apj]
  {10.1088/0004-637X/778/2/165}, \href
  {https://ui.adsabs.harvard.edu/abs/2013ApJ...778..165V} {778, 165}

\bibitem[\protect\citeauthoryear{{Veledina}, {Revnivtsev}, {Durant}, {Gandhi}
  \& {Poutanen}}{{Veledina} et~al.}{2015}]{Veledina_J1753QPO_2015}
{Veledina} A.,  {Revnivtsev} M.~G.,  {Durant} M.,  {Gandhi} P.,   {Poutanen}
  J.,  2015, \mn@doi [\mnras] {10.1093/mnras/stv2201}, \href
  {https://ui.adsabs.harvard.edu/abs/2015MNRAS.454.2855V} {454, 2855}

\bibitem[\protect\citeauthoryear{{Veledina}, {Gandhi}, {Hynes}, {Kajava},
  {Tsygankov}, {Revnivtsev}, {Durant}  \& {Poutanen}}{{Veledina}
  et~al.}{2017}]{veledina_swiftj1753ccfs_2017}
{Veledina} A.,  {Gandhi} P.,  {Hynes} R.,  {Kajava} J. J.~E.,  {Tsygankov}
  S.~S.,  {Revnivtsev} M.~G.,  {Durant} M.,   {Poutanen} J.,  2017, \mn@doi
  [\mnras] {10.1093/mnras/stx1207}, \href
  {https://ui.adsabs.harvard.edu/abs/2017MNRAS.470...48V} {470, 48}

\bibitem[\protect\citeauthoryear{{Veledina} et~al.,}{{Veledina}
  et~al.}{2019}]{Veledina_Polarisation_2019}
{Veledina} et~al., 2019, \mn@doi [A\&A] {10.1051/0004-6361/201834140}, 623, A75

\bibitem[\protect\citeauthoryear{{Venables} \& {Ripley}}{{Venables} \&
  {Ripley}}{2002}]{VenablesRipley_ModernAppliedStatistics_2002}
{Venables} W.~N.,  {Ripley} B.~D.,  2002, {Modern Applied Statistics with S,
  Fourth Edition}.
Springer, \mn@doi{10.1007/978-0-387-21706-2}

\bibitem[\protect\citeauthoryear{{Vincentelli} \& {Casella}}{{Vincentelli} \&
  {Casella}}{2019}]{Vincentelli_SubSecondVariability_2019}
{Vincentelli} F.~M.,  {Casella} P.,  2019, \mn@doi [Astronomische Nachrichten]
  {10.1002/asna.201913617}, \href
  {https://ui.adsabs.harvard.edu/abs/2019AN....340..319V} {340, 319}

\bibitem[\protect\citeauthoryear{{Vincentelli} et~al.,}{{Vincentelli}
  et~al.}{2021}]{Vincentelli_J1535_2021}
{Vincentelli} F.~M.,  et~al., 2021, \mn@doi [\mnras] {10.1093/mnras/stab475},
  \href {https://ui.adsabs.harvard.edu/abs/2021MNRAS.503..614V} {503, 614}

\bibitem[\protect\citeauthoryear{{Whitehurst} \& {King}}{{Whitehurst} \&
  {King}}{1991}]{WhitehurstKing_Superhumps_1991}
{Whitehurst} R.,  {King} A.,  1991, \mn@doi [\mnras] {10.1093/mnras/249.1.25},
  \href {https://ui.adsabs.harvard.edu/abs/1991MNRAS.249...25W} {249, 25}

\bibitem[\protect\citeauthoryear{{Wijnands} \& {van der Klis}}{{Wijnands} \&
  {van der Klis}}{1999}]{WijnandsvanderKlis_PowerSpectra_1999}
{Wijnands} R.,  {van der Klis} M.,  1999, \mn@doi [\apj] {10.1086/306993},
  \href {https://ui.adsabs.harvard.edu/abs/1999ApJ...514..939W} {514, 939}

\bibitem[\protect\citeauthoryear{{Xu}, {Harrison}  \& {Tomsick}}{{Xu}
  et~al.}{2019}]{ATel13025}
{Xu} Y.,  {Harrison} F.,   {Tomsick} J.,  2019, The Astronomer's Telegram,
  \href {https://ui.adsabs.harvard.edu/abs/2019ATel13025....1X} {13025, 1}

\bibitem[\protect\citeauthoryear{{Yu}, {Zhang}, {Yan}, {Wang}  \& {Bai}}{{Yu}
  et~al.}{2018}]{ATel11510}
{Yu} W.,  {Zhang} J.,  {Yan} Z.,  {Wang} X.,   {Bai} J.,  2018, The
  Astronomer's Telegram, \href
  {https://ui.adsabs.harvard.edu/abs/2018ATel11510....1Y} {11510, 1}

\bibitem[\protect\citeauthoryear{{Zampieri}, {Munari}, {Ochner}  \&
  {Manzini}}{{Zampieri} et~al.}{2019}]{ATel12747}
{Zampieri} L.,  {Munari} U.,  {Ochner} P.,   {Manzini} F.,  2019, The
  Astronomer's Telegram, \href
  {https://ui.adsabs.harvard.edu/abs/2019ATel12747....1Z} {12747, 1}

\bibitem[\protect\citeauthoryear{{Zdziarski}, {Dzielak}, {De Marco}, {Szanecki}
   \& {Niedzwiecki}}{{Zdziarski} et~al.}{2021}]{Zdziarski_GeometryJ1820_2021}
{Zdziarski} A.~A.,  {Dzielak} M.~A.,  {De Marco} B.,  {Szanecki} M.,
  {Niedzwiecki} A.,  2021, arXiv e-prints, \href
  {https://ui.adsabs.harvard.edu/abs/2021arXiv210104482Z} {p. arXiv:2101.04482}

\bibitem[\protect\citeauthoryear{{van der Klis}}{{van der
  Klis}}{2000}]{Klis_Millisecond_2000}
{van der Klis} M.,  2000, \mn@doi [ARA\&A] {10.1146/annurev.astro.38.1.717},
  \href {http://adsabs.harvard.edu/abs/2000ARA%26A..38..717V} {38, 717}

\makeatother
\end{thebibliography}

	
	
	
	\appendix
	
	\section{Appendix}
	
	\subsection{CCF Significance Test} \label{sec:CCF_Sims}
	
	Firstly, we analysed the significance of our CCFs. We simulated lightcurves based on our optical data, uncorrelated with X-rays, and then ran CCFs on those. To do this, we Fourier transformed the lightcurves, randomised the phases (i.e. the arguments of the resulting complex numbers), and then inverse Fourier transformed the result (using methodology laid out in \citealt{TimmerKoenig_PowerLawNoise_1995}). This simulated lightcurve therefore had the same power spectrum as the source lightcurve, but was randomised in time and would thus be uncorrelated with respect to X-rays.
	
\begin{figure}
		\includegraphics[width=\columnwidth]{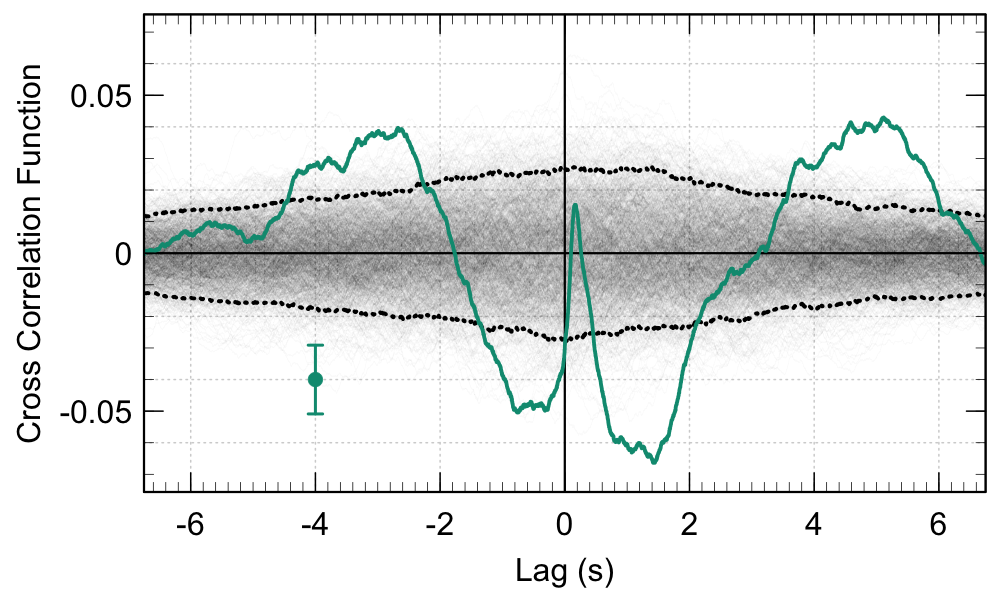}
		\caption{The $g_s$ band vs X-ray CCF from epoch 4, averaged over 162 segments 10\,s in size, is shown in green -- similar to that seen in Figure \ref{fig:10s_CCFs}, but without binning. A representative error bar is plotted. The faded grey area is 1000 overlapping simulated correlation functions, and the black dotted lines are the 5\% and 95\% intervals of all the simulations. The feature at -3\,s, noted in Section \ref{sec:Negative_Lag}, is above this significance line. The -1\,s feature in epoch 6 was similarly found to be significant.}
		\label{fig:CCF_Sims}
	\end{figure}

	This was done 1000 times. Each time, the simulated, uncorrelated lightcurve was cross-correlated with the X-rays, and the resultant CCF was recorded. At the end, for each lag bin, the 5-95\% intervals of all simulations was found. We used this as a way of measuring the significance of features in the original CCF; any features that lie outside of these intervals are considered to be significant. The negative lag feature was found to be outside these intervals for epochs 3--6, and thus we consider it to be a significant feature rather than a spurious result (Fig. \ref{fig:CCF_Sims}).

	\subsection{Simulated Fourier Components} \label{sec:simulated_fourier_components_extra}
	
	As part of our analysis, we wanted to investigate what correlations we would see if we modified certain variability features of the source. We did this by using a custom code that simulated Fourier features, and inputted features akin to those seen in Figs. \ref{fig:powerspectra}--\ref{fig:time_lags}. The code created lightcurves out of these features, and then carrying out cross-correlation analysis on those resultant lightcurves. The majority of the analysis on the simulated lightcurves was carried out by the Stingray\footnote{\url{https://github.com/StingraySoftware/stingray}} python package \citep{Huppenkothen_Stingray_2019}.
	
\begin{figure}
		\includegraphics[width=\columnwidth]{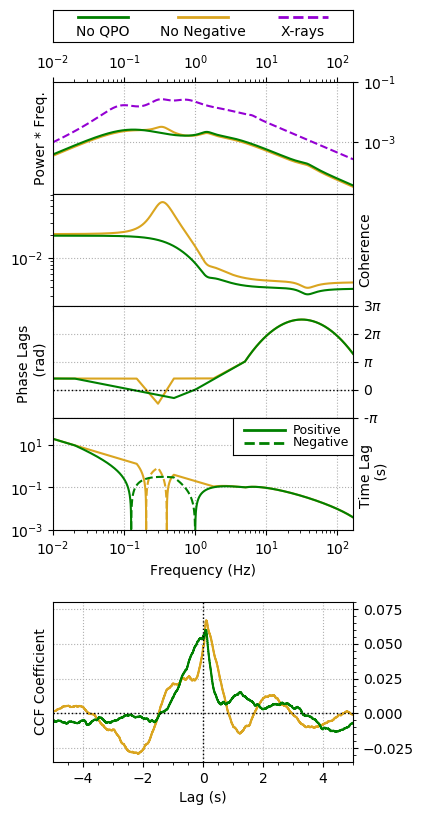}
		\caption{A further two simulations of the $i_{s}$ band with X-rays from epoch 6. \textbf{Top:} Input Fourier components. The green lines are a modification of the behaviour of Epoch 6 with the QPO removed, and the gold lines are a modification that removes the negative lags. \textbf{Bottom:} CCFs made by converting the Fourier components into lightcurves and then cross-correlating the results. CCFs were averaged over multiple 10s segments. Note how each changes affects the CCF in different waves; specifically, the presence of an anti-correlation at -2\,s and a correlation at +3\,s.}
		\label{fig:Epoch6_CorrSim_Extra}
	\end{figure}

	We were thus able to modify the lightcurves by changing the Fourier features. Figure \ref{fig:Epoch6_CorrSim} shows two variants; a simulation meant to reproduce the Epoch 6 features (red) and one that removes all negative-lag trends and the QPO between 0.02--2\,Hz.  Figure \ref{fig:Epoch6_CorrSim_Extra} shows two more variants intended to clarify the contributions of these components -- namely, a version that just removes the QPO (green) and a version that just removes the negative-lag trends (gold).
	
	The synthetic CCFs here reveal which features are due to which Fourier components. The +0.5\,s anti-correlation and the +2\,s correlation, for example, are primarily due to the QPO; while the -1\,s correlation is mainly due to the negative lags. In both, a sub-second lag is still present, showing that it is independent of the variability below 2\,Hz.

	\subsection{Additional Epoch 6 CCFs} \label{sec:epoch6ccfs}
	
	\begin{figure}
		\includegraphics[width=\columnwidth]{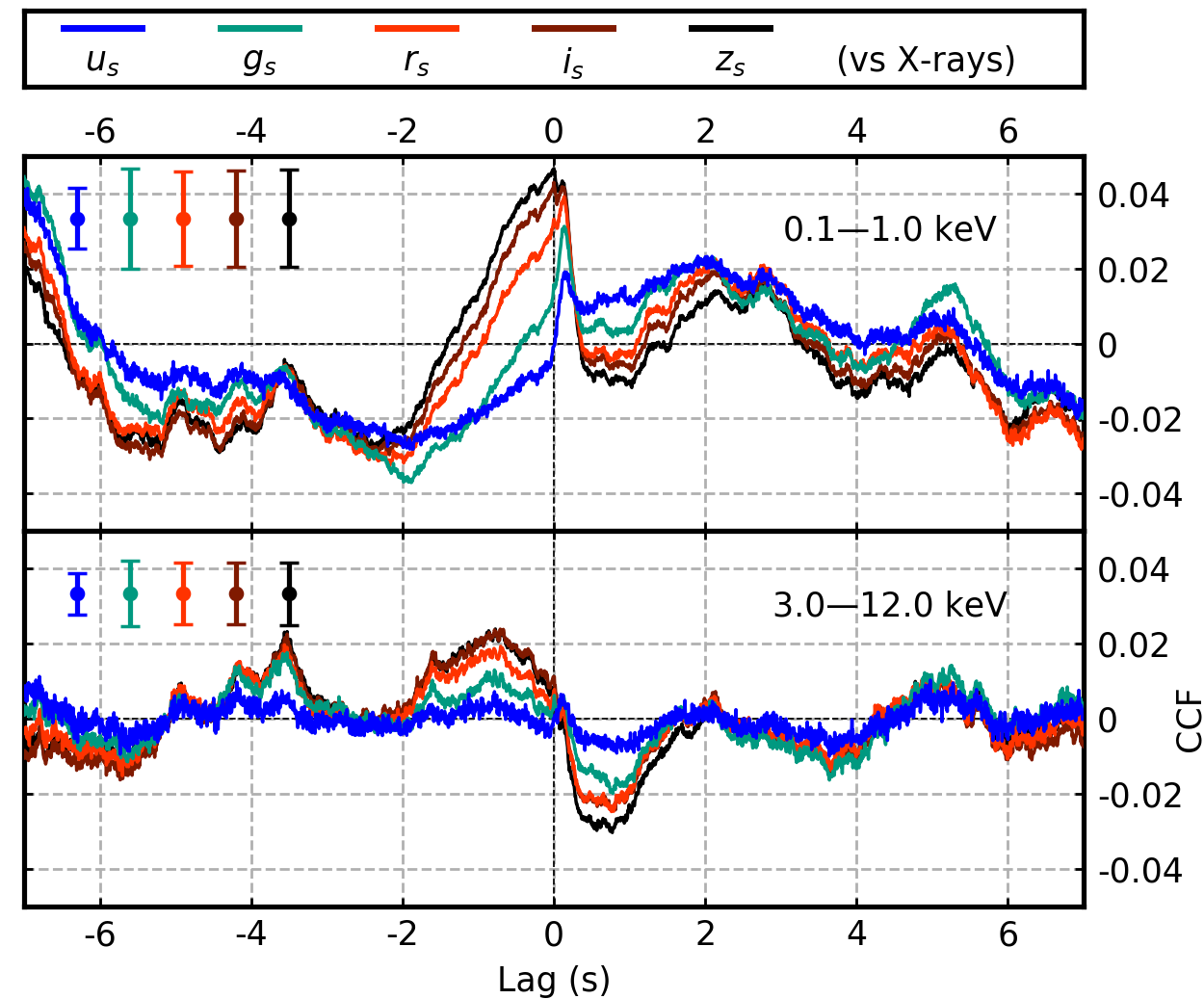}
		\caption{Two additional CCFs from epoch 6, made from 10\,s segments. The CCFs are optical bands vs. soft X-rays (0.1--1.0\,keV, \textbf{Top}) and vs. hard X-rays (3.0--12.0\,keV, \textbf{Bottom}).}
		\label{fig:ccf_xray_ranges}
	\end{figure}
	
   We studied the dependence of the CCF shape on the X-ray energy band for Epoch 6.
   We considered two ranges: $0.1$--$1.0$\,keV and $3.0$--$12.0$\,keV.
   The resulting CCFs are shown in Fig. \ref{fig:ccf_xray_ranges}. 
   The correlations look significantly different in soft and hard X-rays; while the optical/soft-X-ray CCF is dominated by the peak at small negative lags, the shape of the optical/hard X-ray CCF more resembles a sinusoid. 
   Interestingly, the narrow peak at positive lags, which we attribute to the jet, is more prominent in the soft CCF, though is present in both at the same lag.

   Previous studies of the dependence of CCF shape on the X-ray energy band has been performed for Swift~J1753.5--0127 \citep{durant_swift_2008, Durant_OptXCCFs_2011}, GX~339--4, and the neutron star binaries Sco~X-1 and Cyg~X-2 \citep{Durant_OptXCCFs_2011}.
   These show shape variations, yet such acute difference of CCF shape seen in this paper has never been reported before.

   The difference may appear due to presence of two separate components in the X-ray band, e.g. softer component coming from comptonisation of disc photons and harder component coming from hot flow synchrotron Comptonization \citep{veledina_twocomp_2016}.
   Recent spectral studies indeed suggest the presence of two Comptonization continua \citep{Zdziarski_GeometryJ1820_2021}.
   The variability of both of these is caused by the propagating fluctuations, however, their response may alter between harder-when-brighter behaviour for the synchrotron Comptonization to softer-when-brighter for disc Comptonization, resulting in a complex variability pattern \citep{veledina_twocomp_interplay_2018}.
   Correlation with the optical components (disc, jet, and hot flow) is naturally expected to be different for these X-ray components, leading to difference between X-ray energy bands, as the fraction of synchrotron- to disc-Comptonization depends on the energy.
   The weighted average of the sharply different soft and hard CCFs may then lead to the complex CCF shape seen in Fig. \ref{fig:10s_CCFs}.
	
	\subsection{Possibility of a Flared Disc}
	
	\begin{figure}
		\includegraphics[width=\columnwidth]{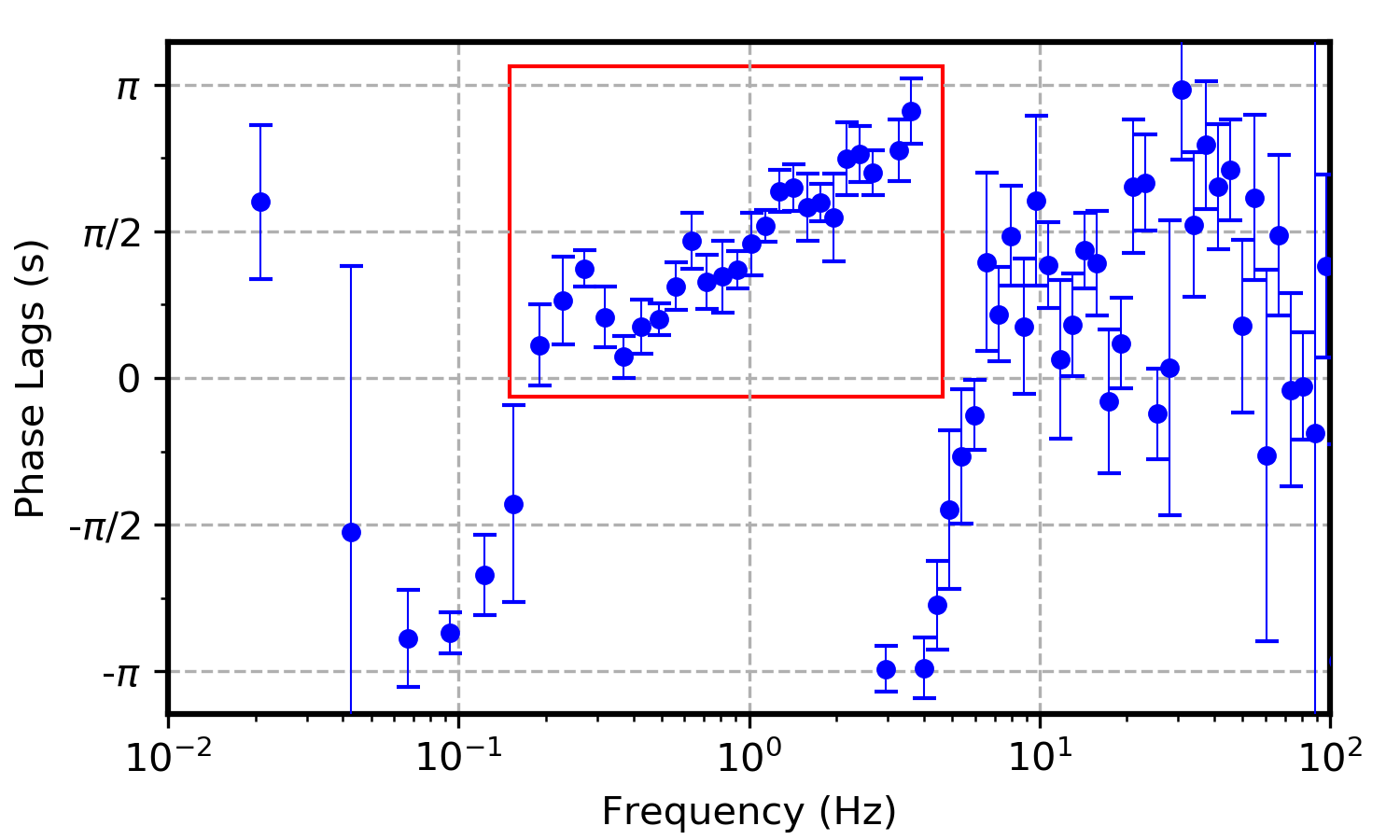}
		\caption{Phase lags from Epoch 4, in the $u_s$ band only, rebinned to best show these variations. The red box indicates a region where there might be a sinusoidal variation.}
		\label{fig:sin_phase_lags}
	\end{figure}
	
	The phase lags for epoch 4 appear to show peaks in the range 0.1--5\,Hz (see Figure \ref{fig:sin_phase_lags}). Such features have previously been proposed to originate in a highly flared disc (\citealt{poutanen_impact_2002} --  see Fig. 6 within). In this model, these peaks can result from features in the power density spectrum, and reflection from the outer disc.
	
	To investigate, we rebinned the data until it best showed this feature. Then, two models were fit to the phase lags over this range, the first being a line, and the second being a line with a sinusoid added on top, both in linear space. However, the second model was not found to fit the phase lags significantly better than the first. There was ambiguity in the results; relative to the variations in the phase lags themselves, large errors are present which allowed for both models to be viable.
	
	Hence, while we do not consider the peaks in epoch 4 to be evidence for a flared disc, we note that this could be a topic for further investigation in future observations of J1820.


	\bsp 
	\label{lastpage}
\end{document}